\documentclass[11pt]{article}
\usepackage{amssymb}
\makeatletter\@addtoreset {equation}{section}\makeatother

\newtheorem{theorem}{Theorem}
\newtheorem{lemma}{Lemma}
\newtheorem{remark}{Remark}
\newtheorem{corollary}{Corollary}

\newtheorem{proposition}{Proposition}

\newenvironment{proof}{
    \noindent {\it Proof.}}{\hfill$\Box$
}

\usepackage[dvips]{epsfig}
\usepackage{graphicx}

\begin{document}

\title{\bf Spectrum of a non-self-adjoint operator \\ associated with
the periodic heat equation}

\author{Marina Chugunova and Dmitry Pelinovsky \\
{\small Department of Mathematics, McMaster University, Hamilton,
Ontario, Canada, L8S 4K1}}

\date{\today}
\maketitle

\begin{abstract}
We study the spectrum of the linear operator $L = -
\partial_{\theta} - \epsilon \partial_{\theta} (\sin \theta \partial_{\theta} )$
subject to the periodic boundary conditions on $\theta \in
[-\pi,\pi]$. We prove that the operator is closed in
$L^2([-\pi,\pi])$ with the domain in $H^1_{\rm per}([-\pi,\pi])$
for $|\epsilon| < 2$, its spectrum consists of an infinite
sequence of isolated eigenvalues and the set of corresponding
eigenfunctions is complete. By using numerical approximations of
eigenvalues and eigenfunctions, we show that all eigenvalues are
simple, located on the imaginary axis and the angle between two
subsequent eigenfunctions tends to zero for larger eigenvalues. As
a result, the complete set of linearly independent eigenfunctions
does not form a basis in $H^1_{\rm per}([-\pi,\pi])$.
\end{abstract}

\section{Introduction}

We address the Cauchy problem for the periodic heat equation
\begin{equation}
\label{heat} \left\{ \begin{array}{ll} \dot{h} = - h_{\theta} -
\epsilon (\sin \theta h_{\theta} )_{\theta}, \quad t > 0,\\
h(0) = h_0,
\end{array} \right.
\end{equation}
subject to the periodic boundary conditions on $\theta \in
[-\pi,\pi]$. This model was derived in the context of the dynamics
of a thin viscous fluid film on the inside surface of a cylinder
rotating around its axis in \cite{Benilov}. Extension of the model
to the three-dimensional motion of the film was reported in
\cite{Benilov2}.

The parameter $\epsilon$ is small for applications in fluid
dynamics \cite{Benilov} and our main results cover the interval
$|\epsilon| < 2$ in accordance to these applications. For any
$\epsilon > 0$, the Cauchy problem for the heat equation
(\ref{heat}) on the half-interval $\theta \in [0,\pi]$ is
generally ill-posed \cite{LS} and it is naturally to expect that
the Cauchy problem remains ill-posed on the entire interval
$\theta \in [-\pi,\pi]$. The authors of the pioneer work
\cite{Benilov} used a heuristic asymptotic solution to suggest
that the growth of "explosive instabilities" might occur in the
time evolution of the Cauchy problem (\ref{heat}).

Nevertheless, in a contradiction with the picture of explosive
instabilities, only purely imaginary eigenvalues were discovered
in the discrete spectrum of the associated linear operator
\begin{equation}
\label{operator-L} L = - \epsilon \frac{\partial}{\partial \theta}
\left( \sin \theta \frac{\partial}{\partial \theta} \right) -
\frac{\partial}{\partial \theta},
\end{equation}
acting on sufficiently smooth periodic functions $f(\theta)$ on
$\theta \in [-\pi, \pi]$. Various approximations of eigenvalues
were obtained in \cite{Benilov} by two asymptotic methods
(expansions in powers of $\epsilon$ and the WKB method) and by
three numerical methods (the Fourier series approximations, the
pseudospectral method, and the Newton--Raphson iterations). The
results of the pseudospectral method were checked independently in
\cite{Trefethen} (see pp. 124--125 and 406--408). It is seen both
in \cite{Benilov} and \cite{Trefethen} that the level sets of the
resolvent of $(\lambda - L)^{-1}$ form divergent curves to the
left and right half-planes and, while true eigenvalues lie on the
imaginary axis, eigenvalues of the truncated Fourier series may
occur in the left and right half-planes of the spectral plane.
This distinctive feature was interpreted in \cite{Benilov} towards
the picture of growth of disturbances and the phenomenon of
explosive instability.

One more question raised in \cite{Benilov} was about the validity
of the series of eigenfunctions associated to the purely imaginary
eigenvalues of the operator $L$ for $\epsilon \neq 0$. Although
various initial conditions $h_0$ were decomposed into a finite sum
of eigenfunctions and the error decreased with a larger number of
terms in the finite sum, the authors of \cite{Benilov} conjectured
that the convergence of the series depended on the time variable
and "even though the series converges at $t = 0$, it may diverge
later". This conjecture would imply that the eigenfunctions of $L$
for $\epsilon \neq 0$ do not form a basis of functions in the
space $H^s([-\pi,\pi])$ with $s > \frac{1}{2}$ unlike the
harmonics of the complex Fourier series associated with the
operator $L$ for $\epsilon = 0$.

In this paper, we prove that the operator $L$ is closed in
$L^2([-\pi,\pi])$ with a domain in $H_{\rm per}^1([-\pi,\pi])$ for
$|\epsilon| < 2$, such that the spectrum of the eigenvalue problem
\begin{equation}
\label{eigenvalue-problem} -\epsilon \frac{d}{d\theta} \left( \sin
\theta \; \frac{d f}{d \theta} \right) - \frac{d f}{ d \theta} =
\lambda f, \qquad f \in H_{\rm per}^1([-\pi,\pi]),
\end{equation}
is well-defined. Here and henceforth, we denote
\begin{equation}
H^1_{\rm per}([-\pi,\pi]) = \left\{ f \in H^1([-\pi,\pi]) : \quad
f(\pi) = f(-\pi) \right\}. \label{H1-per}
\end{equation}
Furthermore, we prove that the residual and essential spectra of
the spectral problem (\ref{eigenvalue-problem}) are empty and the
eigenvalues of the discrete spectrum accumulate at infinity along
the imaginary axis. We further prove completeness of the series of
eigenfunctions associated to all eigenvalues of the discrete
spectrum of $L$ in $H^1_{\rm per}([-\pi,\pi])$. Using the
numerical approximations of eigenvalues and eigenfunctions of the
spectral problem (\ref{eigenvalue-problem}), we show that all
eigenvalues of $L$ are simple, located at the imaginary axis, and
the angle between two subsequent eigenfunctions tends to zero for
larger eigenvalues. As a result, the complete set of linearly
independent eigenfunctions does not form a basis in $H^1_{\rm
per}([-\pi,\pi])$ and hence it cannot be used directly in the
studies of well-posedness of the Cauchy problem associated with
the heat equation (\ref{heat}).

The paper is structured as follows. The domain, closed extensions,
and the essential spectrum of the operator $L$ are analyzed in
Section 2. The discrete spectrum of the operator $L$ is
characterized in Section 3. Section 4 presents numerical
approximations of eigenvalues and eigenfunctions of the spectral
problem (\ref{eigenvalue-problem}). Section 5 gives a summary and
discusses the Cauchy problem for the heat equation (\ref{heat}).
Appendix A reports an extension of the spectral problem
(\ref{eigenvalue-problem}) into a self-adjoint problem in a
weighted $L^2$-space. Appendix B reformulates the eigenvalue
problem (\ref{eigenvalue-problem}) as the resonance pole problem
for a linear Schr\"{o}dinger operator on an infinite line. These
Appendices are not related to our main results and are left in the
text for future work and references.

\section{General properties of the linear operator $L$}

It is obvious that the operator $L$ is densely defined in
$L^2([-\pi,\pi])$ on the space of smooth functions with periodic
boundary conditions. However, the operator $L$ is not closed in
$L^2([-\pi,\pi])$ if the functions are infinitely smooth. We shall
therefore prove that the operator $L$ admits a closure in
$L^2([-\pi,\pi])$ with a domain in $H^1_{\rm per}([-\pi,\pi])$
(Lemmas \ref{lemma-domain} and \ref{lemma-closure}). Properties of
the eigenfunctions and eigenvalues of the spectral problem
(\ref{eigenvalue-problem}) in $H^1_{\rm per}([-\pi,\pi])$ are then
studied (Lemmas \ref{lemma-eigenvalues} and \ref{lemma-symmetry})
and the absence of the residual and essential spectra is
rigorously proved (Lemmas \ref{lemma-residual} and
\ref{lemma-essential}).

\begin{lemma}
If $L f = F$ and $F \in L^2([-\pi,\pi])$ then $ f \in H^1_{\rm
per}([-\pi,\pi]) $ for $|\epsilon| < 2$. \label{lemma-domain}
\end{lemma}

\begin{proof}
For simplicity, we will set $\epsilon > 0$. The case $\epsilon <
0$ is considered similarly. Let $f(\theta)$ be in the domain of
$L$ on $\theta \in [-\pi,\pi]$, such that $L f = F$ for any $F \in
L^2([-\pi,\pi])$. By the method of variation of constants, we
express $f'(\theta)$ from the solution of the first-order ODE for
$f'(\theta)$:
\begin{equation}
\label{integral-equation} f'(\theta) =
\frac{\cot^{1/\epsilon}(\theta/2)}{\sin \theta} \left[
\int_{\theta_0}^{\theta} \tan^{1/\epsilon}(\theta_1/2) F(\theta_1)
d \theta_1 + C \right],
\end{equation}
where $C$ is an integration constant for any $\theta_0 \in
[-\pi,\pi]$. The function $f'(\theta)$ is continuously
differentiable at any $\theta \in (-\pi,0) \cup (0,\pi)$. Let us
consider the behavior of $f'(\theta)$ at the regular singular
points $\theta_0 = 0$ and $\theta_0 = \pi$ on $\theta \in
[0,\pi]$. A similar consideration holds at the regular singular
points $\theta_0 = 0$ and $\theta_0 = - \pi$ on $\theta \in
[-\pi,0]$. For $\theta_0 = 0$ and sufficiently small $\theta
> 0$, it follows from (\ref{integral-equation}) that
$$
\left| \int_{0}^{\theta} \tan^{1/\epsilon}(\theta_1/2) F(\theta_1)
d \theta_1 \right|^2 \leq \int_0^{\theta}
 \tan^{2/\epsilon}(\theta_1/2) d \theta_1 \; \int_0^{\theta}
F^2(\theta_1) d \theta_1 \leq \alpha_1^2(\theta)
\theta^{2/\epsilon + 1},
$$
where $\lim\limits_{\theta \to 0^+} \alpha_1(\theta) = 0$. As a
result, for $\theta > 0$, it holds that
\begin{equation}
\label{singular-point-1} |f'(\theta)| \leq
\frac{\alpha_1(\theta)}{\theta^{1/2}} + \frac{C
\alpha_2(\theta)}{\theta^{1/\epsilon + 1}},
\end{equation}
where $\lim\limits_{\theta \to 0^+} \alpha_2(\theta) \neq 0$. If
$\epsilon < 2$, $f' \in L^2$ near $\theta = 0$ if and only if $C =
0$.

\noindent For $\theta_0 = \pi$ and sufficiently small $\pi -
\theta > 0$, it follows from (\ref{integral-equation}) that
\begin{eqnarray*}
\left| \int_{\pi}^{\theta} \tan^{1/\epsilon}(\theta_1/2)
F(\theta_1) d \theta_1 \right|^2 & \leq & \int_{\theta}^{\pi} (\pi
- \theta_1)^{-2/\epsilon} d \theta_1 \; \int_{\theta}^{\pi} (\pi -
\theta)^{2/\epsilon} \tan^{2/\epsilon}(\theta_1/2) F^2(\theta_1) d
\theta_1 \\ & \leq & \alpha_1^2(\theta) (\pi -
\theta)^{-2/\epsilon + 1},
\end{eqnarray*}
where $\lim\limits_{\theta \to \pi^-} \alpha_1(\theta) = 0$. As a
result, for $\theta < \pi$, it holds that
\begin{equation}
\label{singular-point-2} |f'(\theta)| \leq
\frac{\alpha_1(\theta)}{(\pi - \theta)^{1/2}} + C \alpha_2(\theta)
(\pi - \theta)^{1/\epsilon - 1},
\end{equation}
where $\lim\limits_{\theta \to \pi^-} \alpha_2(\theta) \neq 0$. If
$\epsilon < 2$, $f' \in L^2$ near $\theta = \pi$ for any $C \neq
0$. Finally, we recall the Neumann--Poincare inequality on $\theta
\in [-\pi,\pi]$:
\begin{equation}
\label{NP-inequality} \| f \|^2_{L^2} \leq 4 \pi^2 \| f'
\|^2_{L^2} + \frac{1}{2 \pi} \left( \int_{-\pi}^{\pi} f(\theta) d
\theta \right)^2.
\end{equation}
Due to the estimates (\ref{singular-point-1}) and
(\ref{singular-point-2}), we can see that $\| f \|_{L^1}$ is
bounded on $\theta \in [-\pi,\pi]$ and so is $\| f \|_{L^2}$.
Therefore, the solution $f(\theta)$ of $L f = F$ with $F \in
L^2([-\pi,\pi])$ lies in $H^1([-\pi,\pi])$ if $|\epsilon| < 2$.
\end{proof}

\begin{lemma}
The operator $L$ admits a closure in $L^2([-\pi,\pi])$ for
$|\epsilon| < 2$ with $Dom(L)$ $\subset$ $H^1_{\rm
per}([-\pi,\pi])$. \label{lemma-closure}
\end{lemma}

\begin{proof}
According to Lemma 1.1.2 in \cite{Davies}, if an operator has a
non-empty spectrum in a proper subset of a complex plane, then it
must be closed. The operator $L$ has a non-empty spectrum in
$H^1_{\rm per}([-\pi,\pi])$ since $\lambda = 0$ is an eigenvalue
with the eigenfunction $f_0(\theta) = 1 \in H^1_{\rm
per}([-\pi,\pi])$. We should show that there exists at least one
regular point $\lambda_0 \in \mathbb{C}$, such that
\begin{equation}
\label{regular} \forall f \in H^1_{\rm per}([-\pi,\pi]) : \quad \|
(L - \lambda_0 I) f \|_{L^2} \geq k_0 \| f \|_{L^2}
\end{equation}
for some $k_0 > 0$. In particular, we show that any $\lambda_0 \in
\mathbb{R}$ is a regular point of $L$ in $H_0 \subset H^1_{\rm
per}([-\pi,\pi])$, where
\begin{equation}
H_0 = \left\{ f \in H^1_{\rm per}([-\pi,\pi]) : \quad
\int_{-\pi}^{\pi} f(\theta) d \theta = 0 \right\}.
\end{equation}
By using straightforward computations, we obtain
\begin{equation}
\label{equality-0} (f',Lf) = - \int_{-\pi}^{\pi} \left(1 +
\epsilon \cos \theta \right) |f'|^2 d \theta - \epsilon
\int_{-\pi}^{\pi} \sin \theta \bar{f}' f'' d\theta,
\end{equation}
where $(g,f) = \int_{-\pi}^{\pi} \bar{g}(\theta) f(\theta) d
\theta$ is a standard inner product in $L^2$. If $f \in H^1_{\rm
per}([-\pi,\pi])$, then
\begin{equation}
\label{equality-0a} {\rm Re}(f',f) = 0, \qquad {\rm Re} (f',Lf) =
-\int_{-\pi}^{\pi} \left(1 + \frac{\epsilon}{2} \cos \theta
\right) |f'|^2 d \theta,
\end{equation}
such that for any $\lambda_0 \in \mathbb{R}$ it is true that
$$
|{\rm Re} (f',(L - \lambda_0 I) f) | \geq \left(1 -
\frac{|\epsilon|}{2} \right) \| f' \|_{L^2}^2.
$$
By using the Cauchy--Schwarz inequality, we estimate the
left-hand-side term from above
$$
|{\rm Re}(f',(L - \lambda_0 I) f)| \leq |(f',(L - \lambda_0 I) f)|
\leq \| f'\|_{L^2} \|(L - \lambda_0 I) f \|_{L^2},
$$
such that
\begin{equation}
\label{singular-point-3} \|(L - \lambda_0 I) f\|_{L^2} \geq
\left(1 - \frac{|\epsilon|}{2} \right) \| f' \|_{L^2}.
\end{equation}
Using the Neumann--Poincare inequality (\ref{NP-inequality}) for
any $f \in H_0 \subset H^1_{\rm per}([-\pi,\pi])$, we continue the
right-hand-side of the inequality (\ref{singular-point-3}) and
recover the inequality (\ref{regular}) for any $\lambda_0 \in
\mathbb{R}$ with
$$
k_0 = \frac{1}{2\pi} \left(1 - \frac{|\epsilon|}{2} \right) > 0.
$$
The estimate holds if $|\epsilon| < 2$.
\end{proof}

\begin{remark}
{\rm The formal adjoint of $L$ in $L^2([-\pi,\pi])$ is $L^* =
-\epsilon \partial_{\theta} \left( \sin \theta \partial_{\theta}
\right) + \partial_{\theta}$. According to Lemma 1.2.1 in
\cite{Davies}, the operator $L^*$ also admits a closure in
$L^2([-\pi,\pi])$ with ${\rm Dom}(L^*) \subset H^1_{\rm
per}([-\pi,\pi])$ for $|\epsilon| < 2$.} \label{corollary-closure}
\end{remark}

\begin{lemma}
Let $\lambda$ be a isolated eigenvalue of the spectral problem $L
f = \lambda f$ with an eigenfunction $f \in H_{\rm
per}^1([-\pi,\pi])$. Then,

\begin{itemize}
\item[(i)] $-\lambda$, $\bar{\lambda}$ and $-\bar{\lambda}$ are
also eigenvalues of the spectral problem $L f = \lambda f$ with
the eigenfunctions $f(-\theta)$, $\bar{f}(\theta)$ and
$\bar{f}(-\theta)$ in $H_{\rm per}^1([-\pi,\pi])$.

\item[(ii)] $\lambda$ is also an eigenvalue of the adjoint
spectral problem $L^* f^* = \lambda f^*$ with the eigenfunction
$f^* = f(\pi - \theta)$ in $H_{\rm per}^1([-\pi,\pi])$.

\item[(iii)] $\lambda$ is a simple eigenvalue of $L f = \lambda f$
if and only if $(f^*,f) \neq 0$.
\end{itemize}
\label{lemma-eigenvalues}
\end{lemma}

\begin{proof}
(i) Due to inversion $\theta \to -\theta$, the spectral problem
(\ref{eigenvalue-problem}) transforms to itself with the
transformation $\lambda \to -\lambda$. Due to the complex
conjugation, it transforms to itself with $\lambda \to
\bar{\lambda}$. (ii) Due to the transformation $\theta \to \pi
-\theta$, the spectral problem (\ref{eigenvalue-problem})
transforms to the adjoint problem $L^* f = \lambda f$ with the
same eigenvalue. (iii) The assertion follows by the Fredholm
Alternative Theorem for isolated eigenvalues.
\end{proof}

\begin{lemma}
Let $\lambda$ be an eigenvalue of the spectral problem
(\ref{eigenvalue-problem}) with the eigenfunction $f \in H^1_{\rm
per}([-\pi,\pi])$. Then,
\begin{equation}
\label{equality-3} {\rm Re}(\lambda) = \epsilon \frac{(f',\sin
\theta f')}{(f,f)}, \qquad i {\rm Im}(\lambda) =
\frac{(f',f)}{(f,f)},
\end{equation}
and ${\rm Im}(\lambda) \neq 0$ except for a simple zero eigenvalue
$\lambda = 0$. \label{lemma-symmetry}
\end{lemma}

\begin{proof}
By constructing the quadratic form for $f \in H^1_{\rm
per}([-\pi,\pi])$, we obtain
\begin{equation}
\label{equality-1} (f,Lf) = \epsilon \int_{-\pi}^{\pi} \sin \theta
|f'|^2 d \theta - \int_{-\pi}^{\pi} \bar{f} f' d \theta,
\end{equation}
where the second term is purely imaginary since
\begin{equation}
f \in H^1_{\rm per}([-\pi,\pi]) : \quad \int_{-\pi}^{\pi} \bar{f}'
f d \theta = |f(\theta)|^2 |_{\theta = -\pi}^{\theta = \pi} -
\int_{-\pi}^{\pi} \bar{f} f' d \theta =
-\overline{\int_{-\pi}^{\pi} \bar{f}' f d \theta}.
\end{equation}
Moreover, the equality (\ref{equality-0a}) can be rewritten in the
form
\begin{equation}
\label{equality-minus} i {\rm Im}(\lambda) (f',f) = {\rm Re}
(f',Lf) = - \int_{-\pi}^{\pi} \left( 1 + \frac{\epsilon}{2} \cos
\theta \right) |f'(\theta)|^2 d \theta \leq - \left(1 -
\frac{|\epsilon|}{2} \right) \| f' \|^2_{L^2},
\end{equation}
where the right-hand side is negative if $|\epsilon| < 2$ and
$f(\theta)$ is not constant on $\theta \in [-\pi,\pi]$. Therefore,
$(f',f) \neq 0$ and ${\rm Im}(\lambda) \neq 0$. Finally, the
constant eigenfunction $f(\theta) = 1$ corresponds to the
eigenvalue $\lambda = 0$ and it is a simple eigenvalue since
$(f^*,f) \neq 0$, where $f^*(\theta) = f(\pi-\theta) = 1$ is an
eigenfunction of the adjoint operator $L^*$ for the same
eigenvalue $\lambda = 0$.
\end{proof}

\begin{lemma}
The residual spectrum of the operator $L$ in $L^2([-\pi,\pi])$ is
empty. \label{lemma-residual}
\end{lemma}

\begin{proof}
By a contradiction, assume that $\lambda$ belongs to the residual
part of the spectrum of $L$ such that ${\rm Ker}(L - \lambda I) =
\varnothing$ but ${\rm Range}(L - \lambda I)$ is not dense in
$L^2([-\pi,\pi])$. Let $g \in L^2([-\pi,\pi])$ be orthogonal to
${\rm Range}(L - \lambda I)$, such that
$$
\forall f \in L^2([-\pi,\pi]) : \quad 0 = (g,(L - \lambda I)f) =
((L^* - \bar{\lambda} I) g,f).
$$
Therefore, $(L^* - \bar{\lambda} I) g = 0$, that is
$\bar{\lambda}$ is an eigenvalue of $L^*$.  By Lemma
\ref{lemma-eigenvalues}(ii), $\bar{\lambda}$ is an eigenvalue of
$L$ and by Lemma \ref{lemma-eigenvalues}(i), $\lambda$ is also an
eigenvalue of $L$. Hence $\lambda$ can not be in the residual part
of the spectrum of $L$.
\end{proof}

\begin{lemma}
The essential spectrum of the operator $L$ in $L^2([-\pi,\pi])$ is
empty. \label{lemma-essential}
\end{lemma}

\begin{proof}
According to Theorem 4 on p.1438 in \cite{DS}, if $L$ is a
differential operator defined on the interval $\theta \in
(-\pi,\pi) = (-\pi,0) \cup (0,\pi)$ and $L_{\pm}$ are restrictions
of $L$ on $\theta \in (-\pi,0)$ and $\theta \in (0,\pi)$, then
$\sigma_e(L) = \sigma_e(L_+) \cup \sigma_e(L_-)$, where
$\sigma_e(L)$ denotes the essential spectrum of $L$. By the
symmetry of the two intervals, it is sufficient to prove that the
operator $L_+$ has no essential spectrum on $\theta \in (0,\pi)$
(independently of the boundary conditions at $\theta = 0$ and
$\theta = \pi$). It is also sufficient to carry out the proof for
$\epsilon
> 0$.

\noindent Let us consider the spectral problem
(\ref{eigenvalue-problem}) on $\theta \in [0,\pi]$ and use the
transformation
$$
\cos \theta = \tanh t, \quad \sin \theta = {\rm sech} t, \qquad t
\in \mathbb{R},
$$
such that $\theta \in [0,\pi]$ is mapped to the infinite line $t
\in \mathbb{R}$. Let $f_+(t) = f(\theta)$ on $\theta \in [0,\pi]$.
The function $f_+(t)$ satisfies the spectral problem
\begin{equation}
\label{spectral-infinite} - \epsilon f_+''(t) + f_+'(t) = \lambda
{\rm sech} t \; f_+(t).
\end{equation}
With a transformation $f_+(t) = e^{t/2 \epsilon} g_+(t)$, the
spectral problem (\ref{spectral-infinite}) is written in the
symmetric form
\begin{equation}
\label{3} -\epsilon g_+''(t) + \frac{1}{4 \epsilon} g_+(t) =
\lambda {\rm sech} t \; g_+(t).
\end{equation}
Thus, our operator is extended to a symmetric operator with an
exponentially decaying weight $\rho(t) = {\rm sech}(t)$. According
to Corollary 3 on p. 1437 in \cite{DS}, if $L$ is a symmetric
operator on an open interval $(a,b)$ and $L_0$ is a self-adjoint
extension of $L$ with respect to some boundary conditions at $x =
a$ and $x = b$, then $\sigma_e(L) = \sigma_e(L_0)$. Here $a =
-\infty$, $b = \infty$, and we need to show that the essential
spectrum of the symmetric problem (\ref{3}) is empty in
$L^2(\mathbb{R})$. This follows by Theorem 7 on p.93 in
\cite{Glazman}: since the weight function $\rho(t)$ of the problem
$-y''(t) - \lambda \rho(t) y(t)= 0$ on $t \in \mathbb{R}$ decays
faster than $1/t^2$ as $|t| \to \infty$, the spectrum of $-y''(t)
- \lambda \rho(t) y(t)= 0$ is purely discrete\footnote{Although
the spectral problem (\ref{3}) has an additional term $C y(t)$
with $C > 0$, this term only makes better the inequality (30) on
p.93 in the proof of Theorem 7 of \cite{Glazman}.}.
\end{proof}

\section{The discrete spectrum of the linear operator $L$}

By results of Lemmas \ref{lemma-eigenvalues},
\ref{lemma-symmetry}, \ref{lemma-residual}, and
\ref{lemma-essential}, the spectral problem
(\ref{eigenvalue-problem}) for $|\epsilon| < 2$ may have only two
types of spectral data besides the simple zero eigenvalue: either
pairs of purely imaginary eigenvalues or quartets of symmetric
complex eigenvalues. We shall prove that there exists an infinite
sequence of eigenvalues $\lambda$ which accumulate to infinity
along the imaginary axis (Lemmas \ref{lemma-analytic} and
\ref{lemma-asymptotics}). We further prove completeness of the
eigenfunctions associated to all isolated eigenvalues of the
spectral problem (\ref{eigenvalue-problem}) (Theorem
\ref{lemma-completeness}). Finally, Theorem
\ref{proposition-basis} formulates a sufficient condition for the
set of eigenfunctions of the spectral problem
(\ref{eigenvalue-problem}) not to form a basis in $H^1_{\rm
per}([-\pi,\pi])$.

\begin{lemma}
Let $0 < \epsilon < 2$ and $\epsilon \neq \frac{1}{n}$, $n \in
\mathbb{N}$. For $\lambda \in \mathbb{C}$, the spectral problem
(\ref{eigenvalue-problem}) admits three sets of linearly
independent solutions $\{ f_1(\theta),f_2(\theta)\}$ given by the
Frobenius series
\begin{equation}
\label{frobenius-1} -\pi < \theta < \pi : \quad f_1 = 1 + \sum_{n
\in \mathbb{N}} c_n \theta^n, \quad f_2 = \theta^{-1/\epsilon}
\left( 1 + \sum_{n \in \mathbb{N}} d_n \theta^n \right),
\end{equation}
or
\begin{equation}
\label{frobenius-2} 0 < \pm \theta < \pi : \quad f_1^{\pm} = 1 +
\sum_{n \in \mathbb{N}} a_n^{\pm} (\pi \mp \theta)^n, \quad
f_2^{\pm} = (\pi \mp \theta)^{1/\epsilon} \left( 1 + \sum_{n \in
\mathbb{N}} b_n^{\pm} (\pi \mp \theta)^n \right),
\end{equation}
where all coefficients are uniquely defined. The solution
$f_1(\theta)$ is an analytic function of $\lambda \in \mathbb{C}$
uniformly on $\theta \in [-\pi,\pi]$. \label{lemma-analytic}
\end{lemma}

\begin{proof}
Existence of two linearly independent solutions on $-\pi < \theta
< \pi$ in the form (\ref{frobenius-1}) and on $0 < \pm \theta <
\pi$ in the form (\ref{frobenius-2}) follows by the ODE analysis
near the regular singular points \cite{CL}. The difference between
the two indices of the indicial equation is $\frac{1}{\epsilon}$
and it is non-integer for $\epsilon \neq \frac{1}{n}$, $n \in
\mathbb{N}$\footnote{An additional logarithmic term $\log(\pi -
\theta)$ may need to be included into the Frobeneus series if
$\epsilon = \frac{1}{n}$, $n \in \mathbb{Z}$.}. Since the spectral
problem (\ref{eigenvalue-problem}) depends analytically on
$\lambda$ and the Frobenius series converges absolutely and
uniformly in between two regular singular points, the solution
$f_1(\theta)$ is analytic in $\lambda \in \mathbb{C}$ for any
fixed $\theta \in (-\pi,\pi)$. Due to uniqueness of the solutions
of the ODE (\ref{eigenvalue-problem}), the solution $f_1(\theta)$
can be equivalently represented by the other solutions
\begin{equation}
\label{scattering-representation} f_1(\theta) = A^{\pm}
f_1^{\pm}(\theta) + B^{\pm} f_2^{\pm}(\theta), \quad 0 < \pm
\theta < \pi,
\end{equation}
where $A^{\pm}$ and $B^{\pm}$ are some constants, while the
functions $f_1^{\pm}(\theta)$ and $f_2^{\pm}(\theta)$ are analytic
in $\lambda \in \mathbb{C}$ for any fixed $\pm \theta \in
(0,\pi]$. By matching analytic solutions for any $\pm \theta \in
(0,\pi)$, we find that $A^{\pm}$ and $B^{\pm}$ are analytic
functions of $\lambda \in \mathbb{C}$, the Frobenius series for
$f_1(\theta)$ converges absolutely and uniformly on $\theta \in
[-\pi,\pi]$, and the solution $f_1(\theta)$ is an analytic
function in $\lambda \in \mathbb{C}$ uniformly on $\theta \in
[-\pi,\pi]$.
\end{proof}

\begin{corollary}
There exists an analytic function $F_{\epsilon}(\lambda)$ on ${\rm
Im}\lambda > 0$, roots of which give isolated eigenvalues of the
spectral problem (\ref{eigenvalue-problem}) with the account of
their multiplicity. The only accumulation point of isolated
eigenvalues in the $\lambda$-plane may occur at infinity.
\label{corollary-accumulation}
\end{corollary}

\begin{proof}
The function $f \in H^1([-\pi,\pi])$ satisfies the spectral
problem (\ref{eigenvalue-problem}) if and only if $f(\theta) = C_0
f_1(\theta)$ on $\theta \in [-\pi,\pi]$, where $C_0 = 1$ thanks to
the scaling invariance of homogeneous equations. By using the
representation (\ref{scattering-representation}), we can find that
$A^{\pm} = \lim\limits_{\theta \to \pm  \pi} f_1(\theta)$ are
uniquely defined analytic functions in $\lambda \in \mathbb{C}$.
The function $F_{\epsilon}(\lambda) = A^+ - A^-$ is analytic
function of $\lambda \in \mathbb{C}$ by construction and zeros of
$F_{\epsilon}(\lambda)$ on ${\rm Im} \lambda > 0$ coincide with
the eigenvalues $\lambda$ of the spectral problem
(\ref{eigenvalue-problem}) with the account of their multiplicity.
If $F_{\epsilon}(\lambda_0) = 0$ for some $\lambda_0 \in
\mathbb{C}$, the corresponding eigenfunction $f(\theta)$ lies in
$H^1_{\rm per}([-\pi,\pi])$, i.e. it satisfies the periodic
boundary conditions $f(\pi) = f(-\pi)$. By analytic function
theory, the sequence of roots of $F_{\epsilon}(\lambda)$ can not
accumulate at a finite point on $\lambda \in \mathbb{C}$.
\end{proof}

\begin{remark}
{\rm We will use the method involving the analytic function
$F_{\epsilon}(\lambda)$ on $\lambda \in \mathbb{C}$ for a
numerical shooting method which enables us to approximate
eigenvalues of the spectral problem (\ref{eigenvalue-problem}).
This method involves less computations than the shooting method
described in Appendix C of \cite{Benilov}. Nevertheless, it is
essentially the same shooting method and it uses the ODE analysis
near the regular singular point (Lemma \ref{lemma-analytic}),
which repeats the arguments in Appendix B of \cite{Benilov}. }
\end{remark}

\begin{lemma}
Fix $0 < \epsilon < 2$ and let $\{ \lambda_n \}_{n \in
\mathbb{N}}$ be a set of eigenvalues of the spectral problem
(\ref{eigenvalue-problem}) with ${\rm Im} \lambda_n > 0$, ordered
in the ascending order of $|\lambda_n|$. There exists a finite
number $N \geq 1$, such that for all $n \geq N$, $\lambda_n = i
\omega_n \in i \mathbb{R}_+$ and
\begin{equation}
\label{asymptotic-distribution} \omega_n = C n^2 + {\rm o}(n^2)
\;\; \mbox{as} \;\; n \to \infty,
\end{equation}
for some $C > 0$. \label{lemma-asymptotics}
\end{lemma}

\begin{proof}
We reduce the spectral problem (\ref{eigenvalue-problem}) to two
uncoupled Schro\"{o}dinger equations on an infinite line. Let
$f(\theta)$ be represented on two intervals $\pm \theta \in
[0,\pi]$ by using the transformations
\begin{equation}
\label{Schrodinger-transformations} \cos \theta = {\rm tanh} t,
\qquad \sin \theta = \pm {\rm sech} t,
\end{equation}
where $t \in \mathbb{R}$. Then, the functions $f_{\pm}(t) =
f(\theta)$ on $\pm \theta \in [0,\pi]$ satisfy the uncoupled
spectral problems
\begin{equation}
\label{Schrodinger-two-problems} -\epsilon f_{\pm}''(t) +
f_{\pm}'(t) = \pm \lambda {\rm sech} t \; f_{\pm}(t), \quad t \in
\mathbb{R},
\end{equation}
The normalization condition $f(0) = 1$ is equivalent to the
condition $\lim\limits_{t \to \infty} f_{\pm}(t) = 1$. The
periodic boundary condition $f(\pi) = f(-\pi)$ is equivalent to
the condition $\lim\limits_{t \to -\infty} f_-(t) = \lim\limits_{t
\to -\infty} f_+(t)$. The linear problems
(\ref{Schrodinger-two-problems}) are reformulated as the quadratic
Ricatti equations by using the new variables
\begin{equation}
\label{Ricatti-two-problems} f_{\pm}(t) = e^{\int_{\infty}^t
S_{\pm}(t') dt'} : \quad  S_{\pm} - \epsilon (S_{\pm}' + S_{\pm}^2
) = \pm \lambda {\rm sech} t.
\end{equation}
We choose a negative root of the quadratic equation in the form
\begin{equation}
\label{chain-fraction} S_{\pm}(t) = \frac{1 - \sqrt{1 \mp 4
\epsilon \lambda {\rm sech} t - 4 \epsilon^2 R_{\pm}}}{2
\epsilon}, \qquad R_{\pm} = S_{\pm}'(t).
\end{equation}
The representation (\ref{chain-fraction}) becomes the chain
fraction if the derivative of $S_{\pm}(t)$ is defined recursively
from the same expression (\ref{chain-fraction}). By using the
theory of chain fractions, we claim that $R_{\pm} = {\rm
O}(\sqrt{|\lambda|})$ as $|\lambda| \to \infty$ uniformly on $t
\in \mathbb{R}$. The function $F_{\epsilon}(\lambda)$ of Corollary
\ref{corollary-accumulation} is now expressed by
\begin{equation}
F_{\epsilon}(\lambda) = \lim_{t \to -\infty} \left[ f_+(t) -
f_-(t) \right] = e^{\int_{-\infty}^{\infty} S_+(t) dt} -
e^{\int_{-\infty}^{\infty} S_-(t) dt}.
\end{equation}
Zeros of $F_{\epsilon}(\lambda)$ are equivalent to zeros of the
infinite set of functions
\begin{equation}
G_n(\lambda) = \frac{1}{4 \pi i \epsilon} \int_{-\infty}^{\infty}
\left[ \sqrt{1 + 4 \epsilon \lambda {\rm sech} t - 4 \epsilon^2
R_-(t)} - \sqrt{1 - 4 \epsilon \lambda {\rm sech} t - 4 \epsilon^2
R_+(t)} \right] dt - n, \label{roots-approximate}
\end{equation}
where $n \in \mathbb{N}$. If $R_{\pm}(t) \equiv 0$, the function
$\tilde{G}_n(\omega) = G(i \omega)$, $n \in \mathbb{N}$ is
real-valued and strictly increasing on $\omega \in \mathbb{R}_+$
with $\tilde{G}_n(0) = -n$. By performing asymptotic analysis, we
compute that
\begin{eqnarray}
\nonumber & \phantom{t} & \frac{1}{4 \pi i \epsilon}
\int_{-\infty}^{\infty} \left[ \sqrt{1 + 4 i \epsilon \omega {\rm
sech} t - 4 \epsilon^2 R_-(t)} - \sqrt{1 - 4 i \epsilon \omega
{\rm sech} t - 4 \epsilon^2 R_+(t)} \right] dt \\
\nonumber & = & \frac{1}{\pi i} \int_{-\infty}^{\infty} \frac{2 i
\omega {\rm sech} t + \epsilon (R_+ - R_-)}{\sqrt{1 + 4 i \epsilon
\omega {\rm sech} t - 4 \epsilon^2 R_-(t)} + \sqrt{1 - 4 i
\epsilon \omega {\rm sech} t - 4\epsilon^2 R_+(t)}} dt \\
\label{asymptotic-value-omega} & = & \frac{\sqrt{\omega}}{\sqrt{2
\epsilon} \pi} \int_{-\infty}^{\infty} \frac{dt}{\sqrt{\cosh t}} +
{\rm o}\left(\sqrt{\omega}\right),
\end{eqnarray}
such that $\lim\limits_{\omega \to \infty} \tilde{G}_n(\omega) =
\infty$. Therefore, there exists exactly one root $\omega =
\omega_n$ of $\tilde{G}_n(\omega)$ for each $n$. Since $R_- =
\bar{R}_+$ for $\lambda = i \omega \in i \mathbb{R}$, each simple
root of $\tilde{G}_n(\omega)$ persists for non-zero values of
$R_{\pm}(t) = {\rm O}(\sqrt{\omega})$ uniformly on $t \in
\mathbb{R}$ as $\omega \to \infty$. According to the asymptotic
result (\ref{asymptotic-value-omega}), the roots $\omega_n$ of
$\tilde{G}_n(\omega)$ satisfy the asymptotic distribution
(\ref{asymptotic-distribution}) with $C = \frac{2 \epsilon
\pi^2}{\left(\int_{-\infty}^{\infty} \frac{dt}{\sqrt{\cosh
t}}\right)^2}$.
\end{proof}

\begin{remark}
{\rm Analysis of Lemma \ref{lemma-asymptotics} extends the formal
WKB approach proposed in Section 3 of \cite{Benilov}. In
particular, the equation (\ref{roots-approximate}) with $R_{\pm} =
0$ has been obtained in Eq. (3.11) of \cite{Benilov}. }
\end{remark}

\begin{theorem}
Let $\{ f_n(\theta) \}_{n \in \mathbb{N}}$ be the set of
eigenfunctions corresponding to the set of eigenvalues $\{
\lambda_n \}_{n \in \mathbb{N}}$ in Lemma \ref{lemma-asymptotics}
with ${\rm Im} \lambda_n > 0$. The set of eigenfunctions is
complete in a subspace of $H_0 \subset H^1_{\rm per}([-\pi,\pi])$.
\label{lemma-completeness}
\end{theorem}

\begin{proof}
By Corollary \ref{corollary-accumulation}, eigenvalues of $L$ with
${\rm Im} \lambda > 0$ accumulate to infinity, such that the
matrix operator $M = L^{-1}$ defined on $H_0 \subset H^1_{\rm
per}([-\pi,\pi])$ is compact. By Lemma \ref{lemma-asymptotics},
there are infinitely many isolated eigenvalues and large
eigenvalues are all purely imaginary, such that $|\lambda_n| =
{\rm O}(n^2)$ as $n \to \infty$. These two facts satisfy two
sufficient conditions of the Lidskii's Completeness Theorem.
According to Theorem 6.1 on p. 302 in \cite{GK}, the set of
eigenvectors and generalized eigenvectors of a compact operator
$M$ in a Hilbert space $H$ is complete if there exists $p > 0$
such that
\begin{equation}
\label{lidskii-condition-1} s_n(M) = o(n^{\frac{-1}{p}}), \quad
\mbox{as} \; n \rightarrow \infty,
\end{equation}
where $s_n$ is a singular number of the operator $M$, and the set
\begin{equation}
\label{lidskii-condition-2} W_M = \{ (M f, f) : \quad f \in H,
\quad \| f \|_{L^2} = 1 \}
\end{equation}
lies in a closed angle $\theta_M$ with vertex at $0$ and opening
$\frac{\pi}{p}$.

\noindent Since the singular numbers $s_n$ are eigenvalues of the
positive self-adjoint operator $(M M^*)^{1/2}$ and the eigenvalues
of $L$ grow like ${\rm O}(n^2)$ as $n \to \infty$, we have $s_n(M)
= {\rm O}(n^{-2})$ as $n \to \infty$, such that the first
condition (\ref{lidskii-condition-1}) is verified with $p = 1$.
Since all ${\rm Im} \lambda_n > 0$ for the set of eigenvalues $\{
\lambda_n \}_{n \in \mathbb{N}}$ of Lemma \ref{lemma-asymptotics},
the spectrum of $M$ lies in the lower half plane, such that the
second condition (\ref{lidskii-condition-2}) is also verified with
$p = 1$ ($\theta_M = \pi$).
\end{proof}

\begin{corollary}
The set of eigenfunctions $\{ f_n(\theta) \}_{n \in \mathbb{Z}}$
with $f_0 = 1$ and $f_{-n} = \bar{f}_n$, $\forall n \in
\mathbb{N}$ is complete in $H^1_{\rm per}([-\pi,\pi])$.
\end{corollary}

\begin{remark}
{\rm Due to linear independence of eigenfunctions for distinct
eigenvalues, the set of eigenfunctions $\{ f_n(\theta) \}_{n \in
\mathbb{Z}}$ is also minimal if all eigenvalues are
simple\footnote{By Lemma \ref{lemma-asymptotics}, all eigenvalues
are simple starting with some $n \geq N$.}. If the set
$\{f_n(\theta) \}_{n \in \mathbb{Z}}$ is complete and minimal, any
function $f \in H^1_{\rm per}([-\pi,\pi])$ can be approximated by
a finite linear combination $f_N(\theta) = \sum\limits_{n=-N}^{N}
c_n f_n(\theta)$, such that for any $\varepsilon > 0$, there
exists $N \geq 1$ and the set of coefficients $\{ c_n \}_{-N \leq
n \leq N}$, such that the inequality $\| f - f_N \|_{H^1_{\rm
per}([-\pi,\pi])} < \epsilon$ holds. This approximation does not
imply that the set $\{ f_n(\theta) \}_{n \in \mathbb{Z}}$ forms a
Schauder basis in $H^1_{\rm per}([-\pi,\pi])$, in which case there
would exist a unique series representation $f(\theta) =
\sum\limits_{n \in \mathbb{Z}} c_n f_n(\theta)$ for any $f \in
H^1_{\rm per}([-\pi,\pi])$, such that $\lim\limits_{N \to \infty}
\| f - f_N \|_{H^1_{\rm per}([-\pi,\pi])} = 0$.}
\end{remark}

\begin{theorem}
Let $\{ f_n(\theta) \}_{n \in \mathbb{Z}}$ be a complete and
minimal set of eigenfunctions of the spectral problem
(\ref{eigenvalue-problem}) for the set of eigenvalues $\{
\lambda_n \}_{n \in \mathbb{Z}}$ in Theorem
\ref{lemma-completeness}. The set of eigenfunctions does not form
a basis in Hilbert space $H^1_{\rm per}([-\pi,\pi])$ if
$\lim\limits_{n \to \infty} \cos(\widehat{f_n, f_{n+1}}) = 1$.
\label{proposition-basis}
\end{theorem}

\begin{proof}
The Banach Theorem defines a condition that the complete and
minimal set of eigenfunctions $\{ f_n(\theta) \}_{n \in
\mathbb{Z}}$ forms a basis in Hilbert space $H^1_{\rm
per}([-\pi,\pi])$. According to the Theorem 2 on page 31 in
\cite{JM}, the complete and minimal set of eigenfunctions forms a
basis if and only if $\sup\limits_{N} \|P_N\| < \infty$, where
$P_N$ is the projector of the linear span $\{f_n\}_{-N \leq n \leq
N}$ in the direction of the linear span $\{f_n\}_{|n| \geq N+1}$.

\noindent Since the Hilbert space $H^1_{\rm per}([-\pi,\pi])$ is a
direct sum of the two linear spans above, the norm of the parallel
projector $P_N$ has the geometrical representation $\|P_N\| =
\frac{1}{\sin \alpha_N}$, where $\alpha_N$ is the angle between
the two linear spans \cite{AG}. This implies that the set
$\{f_n(\theta) \}_{n \in \mathbb{Z}}$ is a basis in the Hilbert
space $H^1_{\rm per}([-\pi,\pi])$ if
\begin{equation}
\label{basis} \cos(\widehat{f_n, f_{n+1}}) =
\frac{|(f_n,f_{n+1})|}{\|f_n\| \|f_{n+1}\|} < 1,
\end{equation}
for sufficiently large $n \in \mathbb{Z}$ \cite{IGMK}. In the
other words, the angles between the eigenfunctions should be
uniformly bounded away from zero as $n \to \infty$. If the angle
tends to zero as $n \to \infty$, the set of eigenvectors is not a
basis in the Hilbert space.
\end{proof}

\section{Numerical approximations}

We approximate isolated eigenvalues of the spectral problem
(\ref{eigenvalue-problem}) for $0 < \epsilon < 2$ numerically. In
agreement with numerical results in \cite{Benilov}, we show that
all eigenvalues in the set $\{ \lambda_n \}_{n \in \mathbb{Z}}$
are simple and purely imaginary. Therefore, the set $\{ \lambda_n
\}_{n \in \mathbb{Z}}$ can be ordered in the ascending order, such
that $\lambda_0 = 0$, $\lambda_n = - \lambda_{-n}$, $\forall n \in
\mathbb{N}$, ${\rm Im}\lambda_n < {\rm Im} \lambda_{n+1}$ and
$\lim\limits_{n \to \infty} |\lambda_n| = \infty$. We also show
that the angle between two subsequent eigenfunctions $f_n(\theta)$
and $f_{n+1}(\theta)$ in the set $\{ f_n(\theta) \}_{n \in
\mathbb{Z}}$ tends to zero as $n \to \infty$.

\subsection{Shooting method}

The numerical shooting method is based on the ODE formulation of
the spectral problem (\ref{eigenvalue-problem}). By Lemma
\ref{lemma-analytic} and Corollary \ref{corollary-accumulation},
complex eigenvalues $\lambda \in \mathbb{C}$ are determined by
roots of the analytic function $F_{\epsilon}(\lambda)$ in the
$\lambda$-plane. The number of complex eigenvalues can be computed
with the winding number theory. The number and location of purely
imaginary eigenvalues can be found from real-valued roots of a
scalar real-valued function.

\begin{proposition}
Let the eigenfunction $f(\theta)$ of the spectral problem
(\ref{eigenvalue-problem}) for $0 < \epsilon < 2$ be normalized by
the condition $f(0) = 1$. The eigenvalue $\lambda$ is purely
imaginary if and only if $f(\theta) = \overline{f}(-\theta)$ on
$\theta \in [-\pi,\pi]$. \label{lemma-reflection}
\end{proposition}

\begin{proof}
If $\lambda \in i \mathbb{R}$ and $f(\theta)$ satisfies the
second-order ODE (\ref{eigenvalue-problem}) on $\theta \in
[-\pi,\pi]$, then $\bar{f}(-\theta)$ satisfies the same ODE
(\ref{eigenvalue-problem}) on $\theta \in [-\pi,\pi]$. By
Corollary \ref{corollary-accumulation}, if $f \in H^1_{\rm
per}([-\pi,\pi])$, $f(0) = 1$ and $0 < \epsilon < 2$, the solution
$f(\theta)$ is uniquely defined. By uniqueness of solutions,
$f(\theta) = \overline{f}(-\theta)$ on $\theta \in [-\pi,\pi]$.

\noindent If $f(\theta) = \overline{f}(-\theta)$ on $\theta \in
[-\pi,\pi]$, then,
$$
\int_{-\pi}^{\pi} \sin \theta |f'(\theta)|^2 d \theta =
\int_{0}^{\pi} \sin \theta |f'(\theta)|^2 d \theta - \int_0^{\pi}
\sin \theta |f'(-\theta)|^2 d \theta = 0,
$$
such that ${\rm Re} \lambda = 0$ according to the equality
(\ref{equality-3}) in Lemma \ref{lemma-symmetry}.
\end{proof}

\begin{corollary}
Let $f(\theta)$ be an eigenfunction of the spectral problem
(\ref{eigenvalue-problem}) for $\lambda \in i \mathbb{R}$, such
that $f \in H^1_{\rm per}([-\pi,\pi])$ and $f(0) = 1$. Then,
$f(\pi) = f(-\pi)$ is equivalent to $f(\pi) \in \mathbb{R}$. The
eigenvalue $\lambda \in i \mathbb{R}$ is simple if and only if
\begin{equation}
\label{Fredholm-constraint} (f^*,f) = 2 {\rm Re} \int_0^{\pi}
f(\theta) \bar{f}(\pi - \theta)  d\theta \neq 0.
\end{equation}
\label{corollary-symmetry}
\end{corollary}

\begin{proof}
The first assertion follows by the symmetry relation $f(\theta) =
\bar{f}(-\theta)$ evaluated at $\theta = \pi$. The second asserion
follows by Lemma \ref{lemma-eigenvalues} with the use of the
symmetry $f^*(\theta) = f(\pi - \theta)$.
\end{proof}

\noindent {\bf Numerical Method:} By using Lemma
\ref{lemma-analytic}, the function $f(\theta)$ with $f(0) = 1$ is
represented uniquely by the Frobenius series
\begin{equation}
\label{Frob-series} f(\theta) = f_1(\theta) = 1 + \sum_{n \in
\mathbb{N}} c_n \theta^n,
\end{equation}
where the coefficients $\{ c_n \}_{n \in \mathbb{N}}$ are uniquely
defined by the recursion relation
\begin{equation}
c_n = - \frac{1}{n (1 + \epsilon n)} \left( \lambda c_{n-1} +
\epsilon n \sum_{m \in \mathbb{N}'} \frac{(-1)^{\frac{n-m}{2}}
m}{(n-m+1)!} c_m \right), \quad n \in \mathbb{N},
\end{equation}
where $c_0 = 1$ and $\mathbb{N}'$ is a set of integers in the
interval $[1,n-2]$ such that $n-m$ is even. For instance,
$$
c_1 = -\frac{\lambda}{1 + \epsilon}, \quad c_2 =
\frac{\lambda^2}{2(1+\epsilon)(1+2\epsilon)}, \quad  c_3 =
-\frac{\lambda (\lambda^2 + \epsilon (1 + 2 \epsilon))}{3!
(1+\epsilon)(1+2\epsilon)(1 + 3 \epsilon)},
$$
and so on. We truncate the power series expansion on $N = 100$
terms and approximate the initial value
$[f(\theta_0),f'(\theta_0)]$ at $\theta_0 = 10^{-8}$. By using the
fourth-order Runge--Kutta ODE solver with time step $h = 10^{-4}$,
we obtain a numerical approximation of $f \equiv f_+(\theta)$ on
$\theta \in [\theta_0, \pi-\theta_0]$ for $\lambda$ and $f \equiv
f_-(\theta)$ on the same interval for $-\lambda$. By Lemma
\ref{lemma-eigenvalues}(i), the numerical approximation of the
function $F_{\epsilon}(\lambda)$ of Corollary
\ref{corollary-accumulation} is
\begin{equation}
\label{F-approximation} \hat{F}_{\epsilon}(\lambda) = f_+(\pi -
\theta_0) - f_-(\pi - \theta_0).
\end{equation}
If $\lambda \in i \mathbb{R}$, the function
$\hat{F}_{\epsilon}(\lambda)$ is simplified by using Corollary
\ref{corollary-symmetry} as $\hat{F}_{\epsilon}(\lambda) = 2i {\rm
Im} f_+(\pi - \theta_0)$. Table 1 represents the numerical
approximations of the first four non-zero eigenvalues $\lambda \in
i \mathbb{R}$ for $\epsilon = 0.5, 1.0, 1.5$\footnote{We note that
the Frobenius series (\ref{Frob-series}) is not affected by the
logarithmic terms for $\epsilon = 0.5$ and $\epsilon = 1.0$, since
$0$ is the largest index of the indicial equation at $\theta =
0$.} with the error computed from the residual
$$
R = \left|\frac{(f,Lf)}{(f,f)} - \lambda\right|.
$$
We can see from Table 1 that the accuracy drops with larger values
of $\epsilon$ and for larger eigenvalues, but the eigenvalues
persist inside the interval $|\epsilon| < 2$.

Figure 1 shows the profiles of eigenfunctions $f(\theta)$ on
$\theta \in [0,\pi]$ for the first two eigenvalues $\lambda = i
\omega_{1,2} \in i \mathbb{R}_+$ for $\epsilon = 0.5$ (left) and
$\epsilon = 1.5$ (right). We can see from Fig. 1 that the
derivative of $f(\theta)$ becomes singular as $\theta \to \pi^-$
for $\epsilon \geq 1$. We can also see that the real part of the
eigenfunction $f(\theta)$ has one zero on $\theta \in (0,\pi)$ for
the first eigenvalue and two zeros for the second eigenvalue,
while the imaginary part of the eigenfunction $f(\theta)$ has a
fewer number of zeros by one. The numerical approximations of the
eigenvalue and eigenfunctions of the spectral problem
(\ref{eigenvalue-problem}) are structurally stable with respect to
variations in $\theta_0$, $N$ and $h$.

Figure 2 shows the complex plane of $w =
\hat{F}_{\epsilon}(\lambda)$ (left) and the argument of $w$
(right) when $\lambda$ traverses along the first quadrant of the
complex plane $\lambda \in \Lambda_1 \cup \Lambda_2 \cup
\Lambda_3$ for $\epsilon = 0.5$. Here $\Lambda_1 = x + i r$ with
$x \in [r,R]$, $\Lambda_2 = R e^{i \varphi}$ with $\varphi \in
[\varphi_0,\frac{\pi}{2}-\varphi_0]$ and $\lambda_3 = r + i y$
with $y \in [r,R]$, where $r = 0.1$, $R = 10$, and $\varphi_0 =
{\rm arctan}(r/R)$. It is obvious that the winding number of
$\hat{F}_{\epsilon}(\lambda)$ across the closed contour is zero.
Therefore, no zeros of $\hat{F}_{\epsilon}(\lambda)$ occurs in the
first quadrant of the complex plane $\lambda \in \mathbb{C}$. The
numerical result is structurally stable with respect to variations
in $r$, $R$ and $\epsilon$.

\begin{center}
\begin{tabular}{|c|l|l|l|l|l|l|l|l|}
\hline $\epsilon$ & $\omega_1$ & $R_1$ & $\omega_2$ & $R_2$ & $\omega_3$ & $R_3$ & $\omega_4$ & $R_4$ \\
\hline $0.5$  & $1.167342$   & $0.000051$ & $2.968852$  & $0.000405$ & $5.483680$  & $0.001436$ & $8.715534$  & $0.003653$ \\
\hline $1.0$  & $1.449323$   & $0.000837$ & $4.319645$  & $0.007069$ & $8.631474$  & $0.024964$ & $14.382886$  & $0.061881$ \\
\hline $1.5$  & $1.757278$   & $0.002691$ & $5.719671$  & $0.018412$ & $11.846709$  & $0.054271$ & $20.138824$  & $0.113834$ \\
\hline
\end{tabular}
\end{center}

{\bf Table 1:} Numerical approximations of the first four
eigenvalues $\lambda = i \omega_n$ of the spectral problem
(\ref{eigenvalue-problem}) and the residuals $R = R_n$ for three
values of $\epsilon$.

\begin{figure}
\begin{center}
\includegraphics[height=5.5cm]{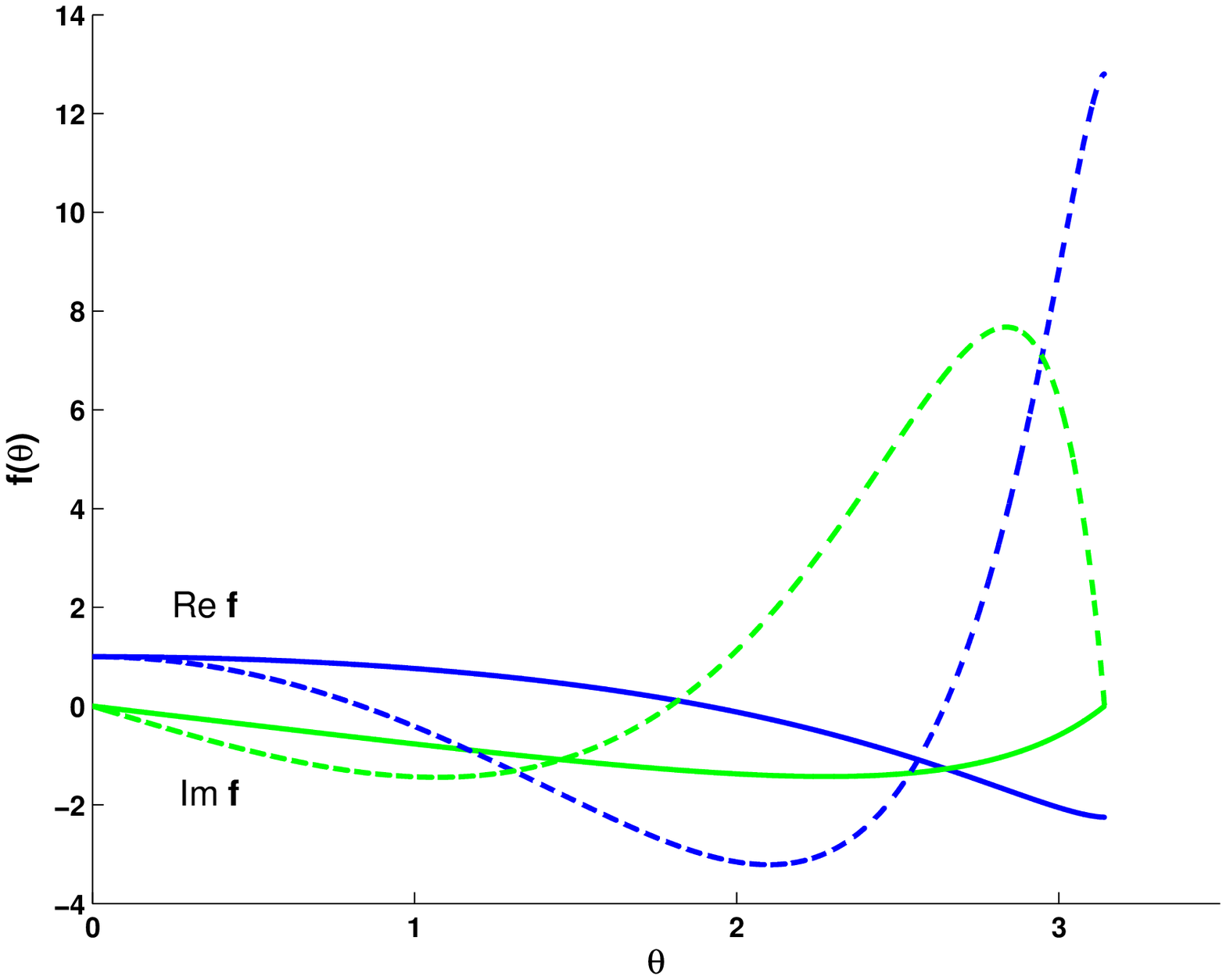}
\includegraphics[height=5.5cm]{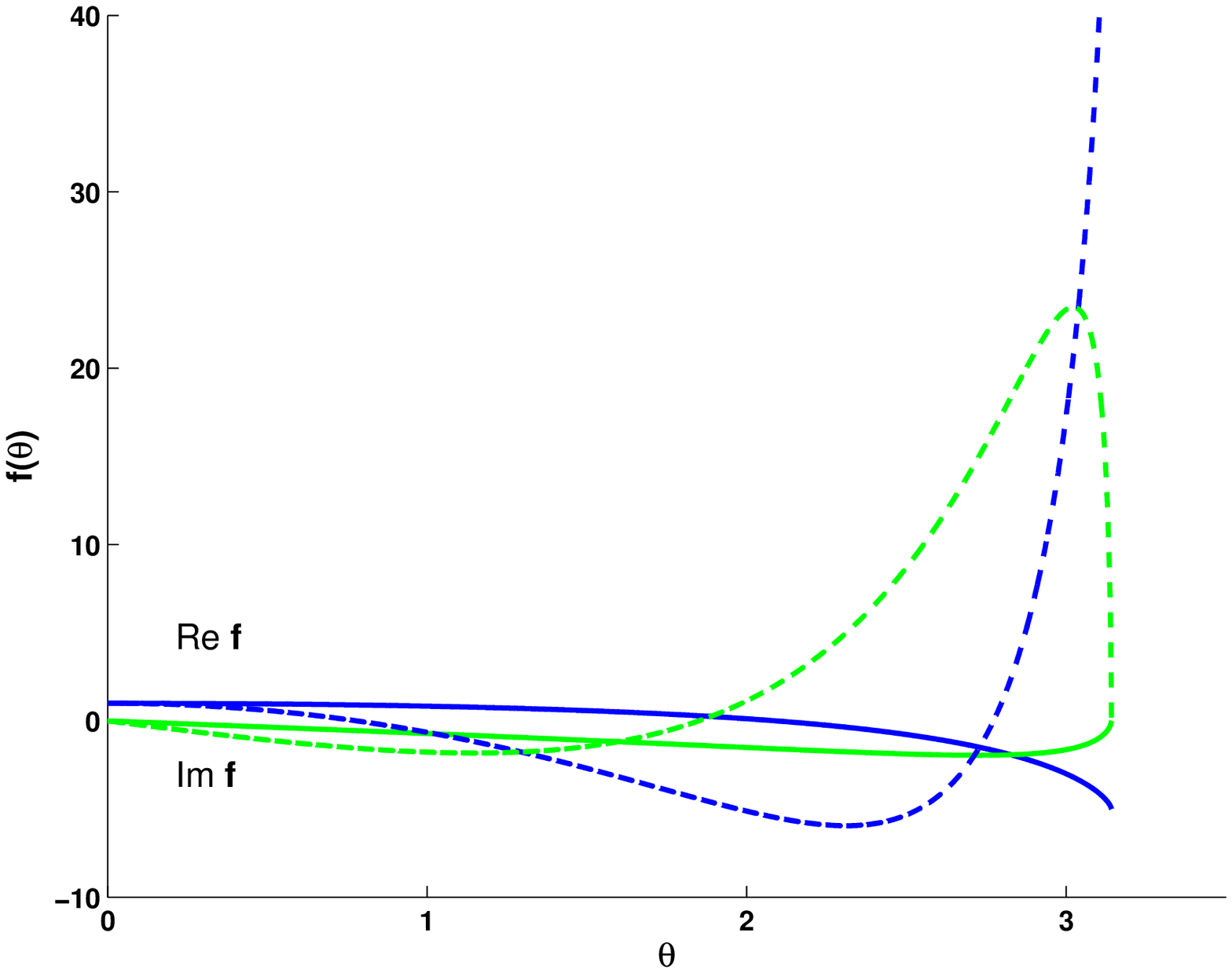}
\end{center}
\caption{The real part (blue) and imaginary part (green) of the
eigenfunction $f(\theta)$ on $\theta \in [0,\pi]$ for the first
(solid) and second (dashed) eigenvalues $\lambda = i \omega_{1,2}
\in i \mathbb{R}_+$ for $\epsilon = 0.5$ (left) and $\epsilon =
1.5$ (right).} \label{fig1}
\end{figure}

\begin{figure}
\begin{center}
\includegraphics[height=5.5cm]{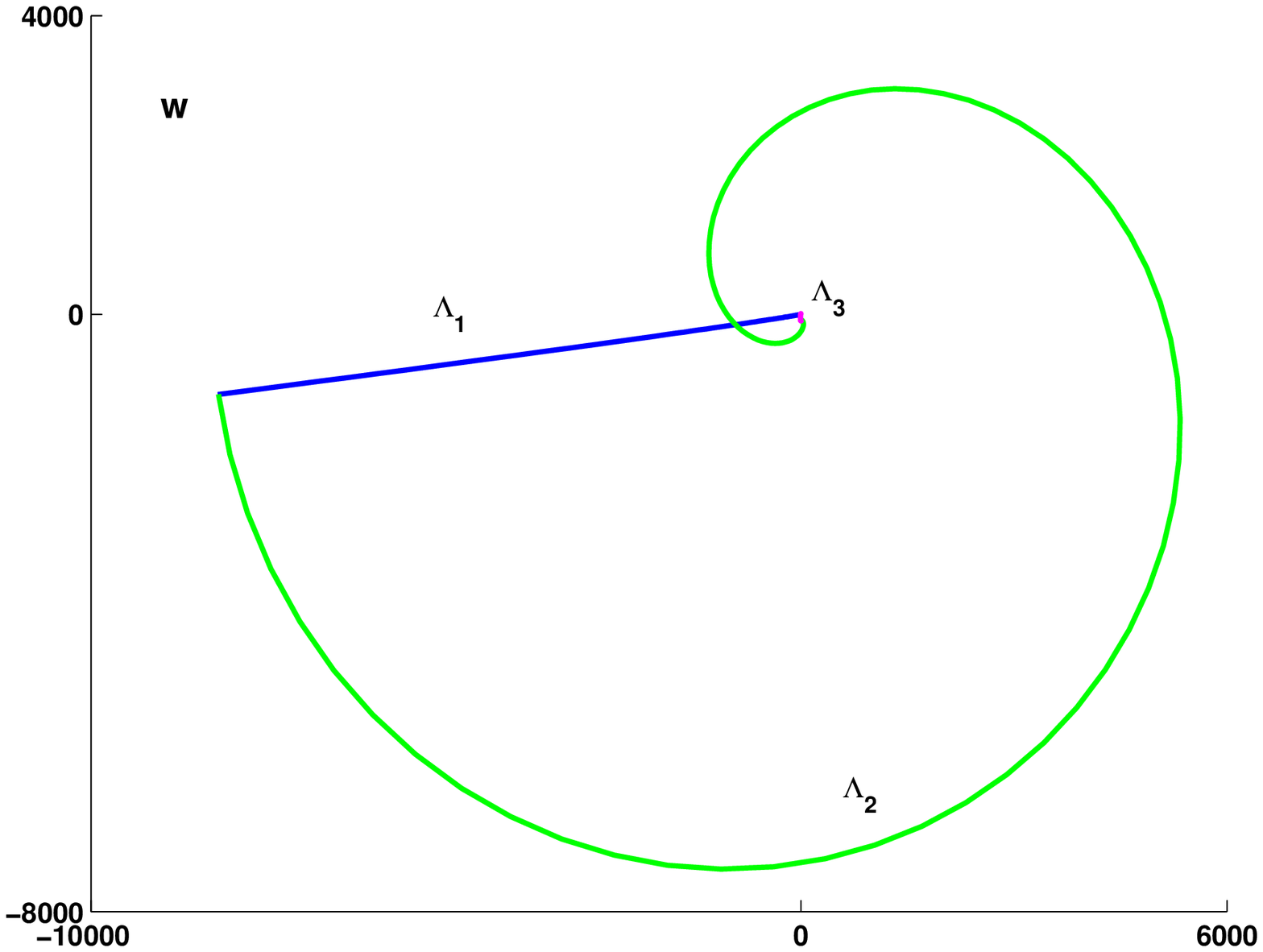}
\includegraphics[height=5.5cm]{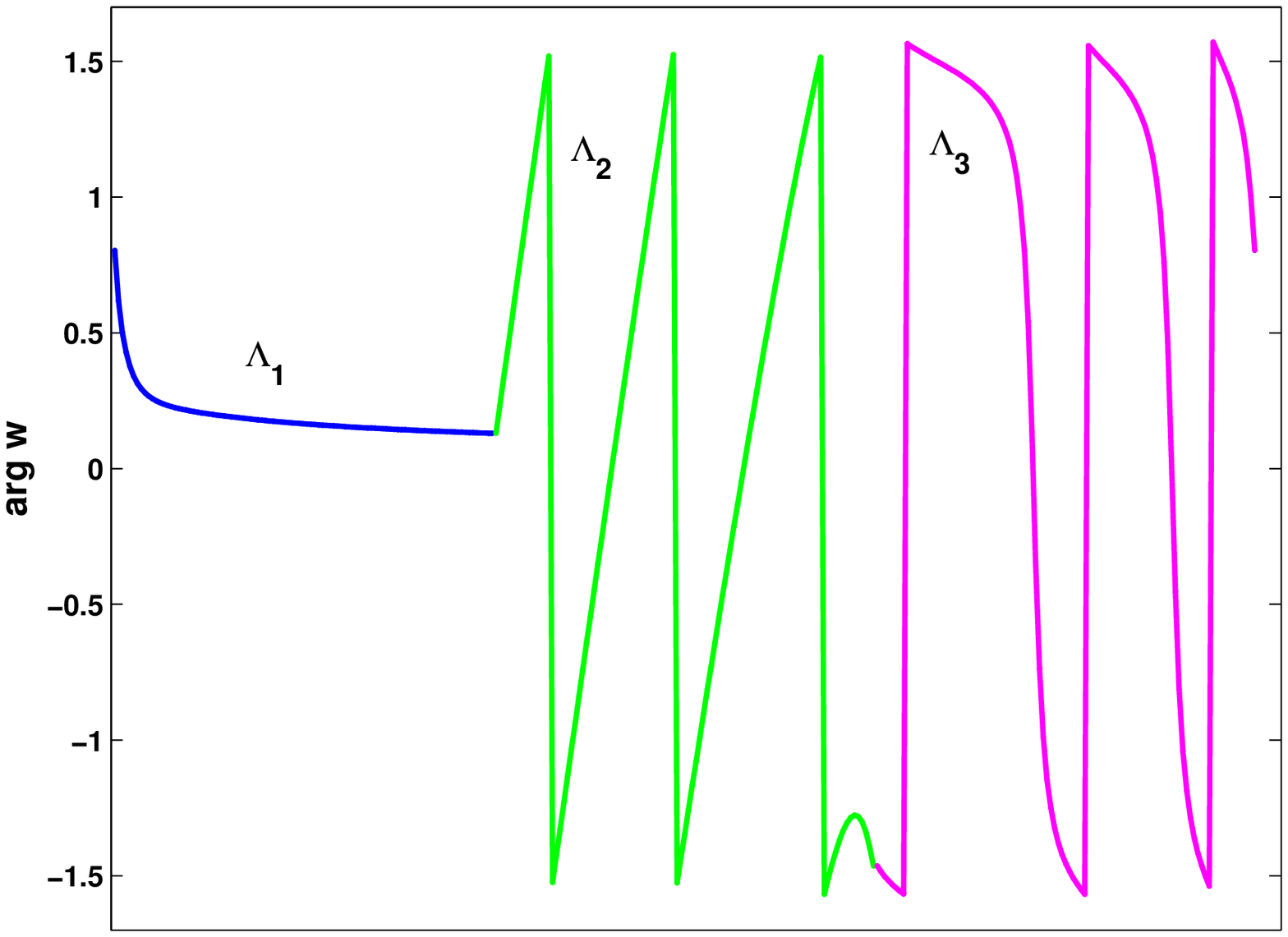}
\end{center}
\caption{The image of the curve $w = \hat{F}_{\epsilon}(\lambda)$,
when $\lambda$ traverses along the contours $\Lambda_1$ (blue),
$\Lambda_2$ (green) and $\Lambda_3$ (magenta) for $\epsilon =
0.5$: the image curve on the $w$-plane (left) and the argument of
$w$ (right).} \label{fig2}
\end{figure}

\subsection{Spectral method}

The numerical spectral method is based on the reformulation of the
second-order ODE (\ref{eigenvalue-problem}) as the second-order
difference equation and on the subsequent truncation of the
difference eigenvalue problem. It is found in \cite{Trefethen}
that the truncation procedure lead to spurious complex eigenvalues
which bifurcate off the imaginary axis.

\noindent {\bf Numerical method:} Let $f(\theta)$ be an
eigenfunction of the spectral problem (\ref{eigenvalue-problem})
in $H^1_{\rm per}([-\pi,\pi])$. This eigenfunction is equivalently
represented by the Fourier series
\begin{equation}
\label{Fourier-series} f(\theta) = \sum_{n \in \mathbb{Z}} f_n
e^{-i n \theta}, \qquad f_n = \frac{1}{2\pi} \int_{-\pi}^{\pi}
f(\theta) e^{i n \theta} d \theta,
\end{equation}
where the infinite-dimensional vector ${\bf f} =
(...,f_{-2},f_{-1},f_0,f_1,f_2,...)$ is defined in ${\bf f} \in
l^2_1(\mathbb{Z})$ equipped with the norm $\| {\bf f} \|^2_{l^2_1}
= \sum_{n \in \mathbb{Z}} (1 + n^2) |f_n|^2 < \infty$. The
spectral problem (\ref{eigenvalue-problem}) for $|\epsilon| < 2$
is equivalent to the difference eigenvalue problem
\begin{equation}
\label{difference-problem} n f_n + \frac{\epsilon}{2}  n \left[
(n+1) f_{n+1} - (n-1) f_{n-1}) \right] = - i \lambda f_n, \qquad n
\in \mathbb{Z}.
\end{equation}
The difference eigenvalue problem (\ref{difference-problem})
splits into three parts
\begin{equation}
\label{4} A {\bf f}_+ = - i\lambda {\bf f}_+, \qquad A {\bf f}_- =
i \lambda {\bf f}_-, \qquad \lambda f_0 = 0,
\end{equation}
where ${\bf f}_{\pm} = (f_{\pm 1}, f_{\pm 2}, ...)$ and $A$ is an
infinite-dimensional matrix
\begin{equation}
\label{matrix-A}
A = \left[ \begin{array}{ccccc} 1 & \epsilon & 0 & 0 & \cdots \\
-\epsilon & 2 & 3 \epsilon & 0 & \cdots \\ 0 & -3 \epsilon & 3 &
6 \epsilon & \cdots \\ 0 & 0 & -6 \epsilon & 4 & \cdots \\
\vdots & \vdots & \vdots & \vdots & \ddots \end{array} \right]
\end{equation}
Since $A = D - i S$, where $D$ is a diagonal matrix and $S$ is a
self-adjoint tri-diagonal matrix, one can define the discrete
counterpart of Lemma \ref{lemma-symmetry}
$$
{\rm Im} \lambda  = \frac{({\bf f}_+,D{\bf f}_+)}{({\bf f}_+,{\bf
f}_+)} = \frac{\sum_{n \in \mathbb{N}} n |f_n|^2}{\sum_{n \in
\mathbb{N}} |f_n|^2}, \qquad {\rm Re} \lambda = \frac{({\bf f}_+,S
{\bf f}_+)}{({\bf f}_+,{\bf f}_+)}.
$$
where ${\rm Im} \lambda > 0$. The adjoint eigenfunction
$f^*(\theta) = f(\pi - \theta)$ is recovered from the eigenvector
${\bf f}$ by ${\bf f}^* = J {\bf f}$, where
$$
J = \left[ \begin{array}{ccc} 0 & 0 & J_0 \\ 0 & 1 & 0 \\ J_0 & 0
& 0 \end{array} \right]
$$
and $J_0$ is a diagonal operator with entries $(-1,1,-1,1,...)$.

According to Theorem \ref{lemma-completeness}, rewritten from the
set of eigenfunctions $\{ f_n(\theta) \}_{n \in \mathbb{Z}}$ to
the set of eigenvectors $\{ {\bf f}_n \}_{n \in \mathbb{Z}}$, the
inverse matrix operator $A^{-1}$ is of the Hilbert-Schmidt type.
Let $A_N^{-1}$ denote the truncation of the matrix operator
$A^{-1}$ at the first $N$ rows and columns. If a sequence of
truncated operators $A_N^{-1}$ converges uniformly to the limiting
compact operator $A^{-1}$ as $N \to \infty$, then the spectra of
matrices $A^{-1}_N$ also converge to the spectrum of $A^{-1}$ as
$N \to \infty$. However, the distance between eigenvalues of
$A_N^{-1}$ and $A^{-1}$ may not be small for fixed $N$.

\begin{figure}
\begin{center}
\includegraphics[height= 6cm]{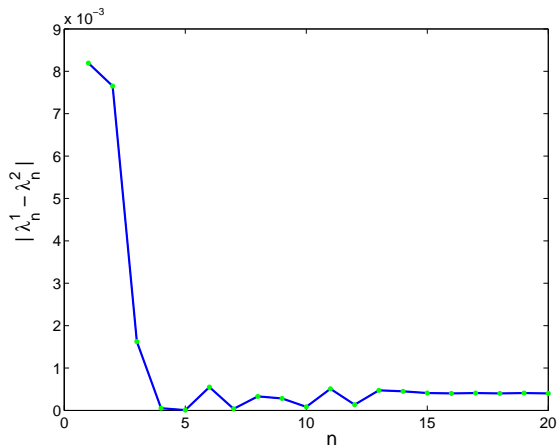}
\end{center}
\caption{The distance between eigenvalues computed by the shooting
and spectral methods for $\epsilon = 0.1$.} \label{fig-dif}
\end{figure}

The smallest eigenvalues of the truncated matrix $A_N^{-1}$ are
found with the parallel Krylov subspace iteration algorithm
\cite{SLEPc}. Figure \ref{fig-dif} shows the distance between
eigenvalues of the shooting method and eigenvalues of the Krylov
spectral method for $\epsilon = 0.1$. The difference between two
eigenvalues is small of the order ${\rm O}(10^{-3})$ but the
advantage of the parallel algorithm is that the calculating time
of 20 largest eigenvalues of $A_N^{-1}$ for $N = 10^6$ takes less
than one minute on a network of 16 processors while finding the
same set of eigenvalues by the shooting method with the time step
$h = 10^{-5}$ takes about one hour.

\begin{figure}
\begin{center}
\includegraphics[height=5.5cm]{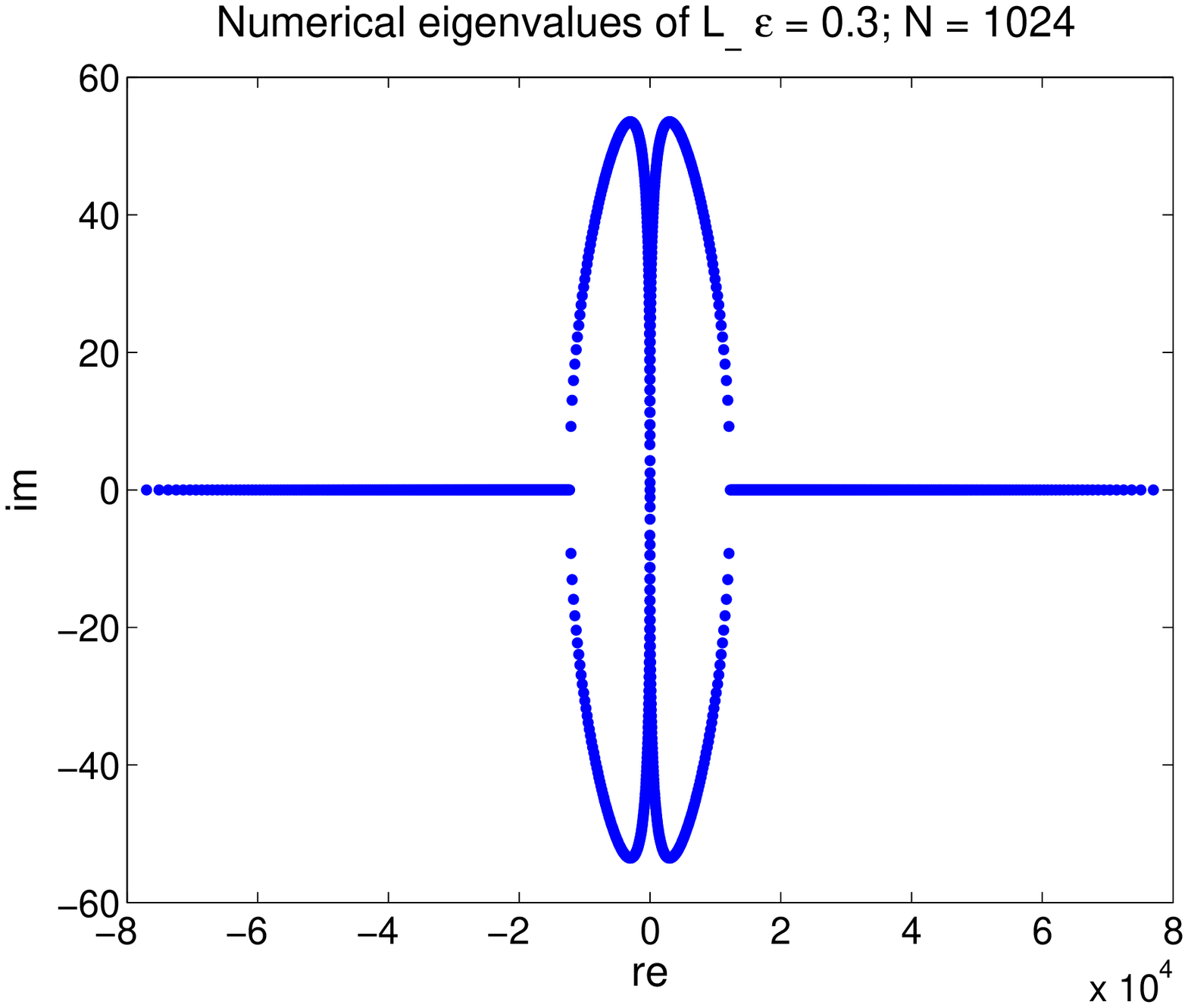}
\includegraphics[height=5.5cm]{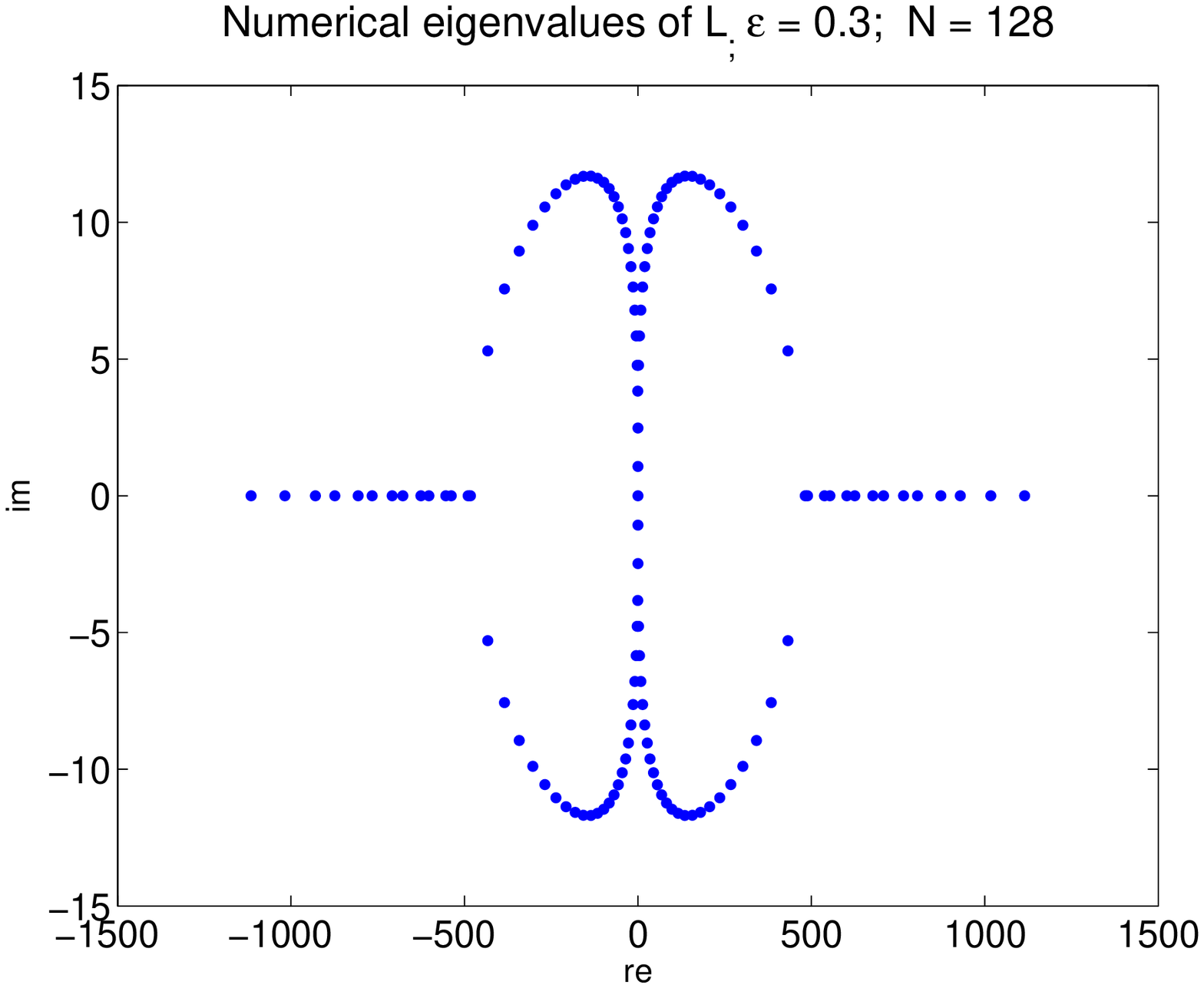}
\end{center}
\caption{Spectrum of the truncated difference eigenvalue problem
(\ref{difference-problem}) for $\epsilon = 0.3$: $N = 128$ (left)
and $N = 1024$ (right).} \label{fig9}
\end{figure}

Figure \ref{fig9} shows symmetric pairs of eigenvalues of the
matrix $A_N$ for $\epsilon = 0.3$ at $N = 128$ (left) and $N =
1024$ (right). We confirm the numerical result of \cite{Trefethen}
that the truncation of the matrix operator $A$ always produces
splitting of large eigenvalues off the imaginary axis. Moreover,
starting with some number $n$, the eigenvalues of $A_N$ are
real-valued. This feature is an artifact of the truncation, which
contradicts to Lemmas \ref{lemma-symmetry} and
\ref{lemma-asymptotics} as well as to results of the shooting
method. However, the larger is $N$, the more eigenvalues remain on
the purely imaginary axis. Therefore, the corresponding
eigenvectors can be used to compute the angle in Theorem
\ref{proposition-basis}.

\begin{figure}
\begin{center}
\includegraphics[height= 6cm]{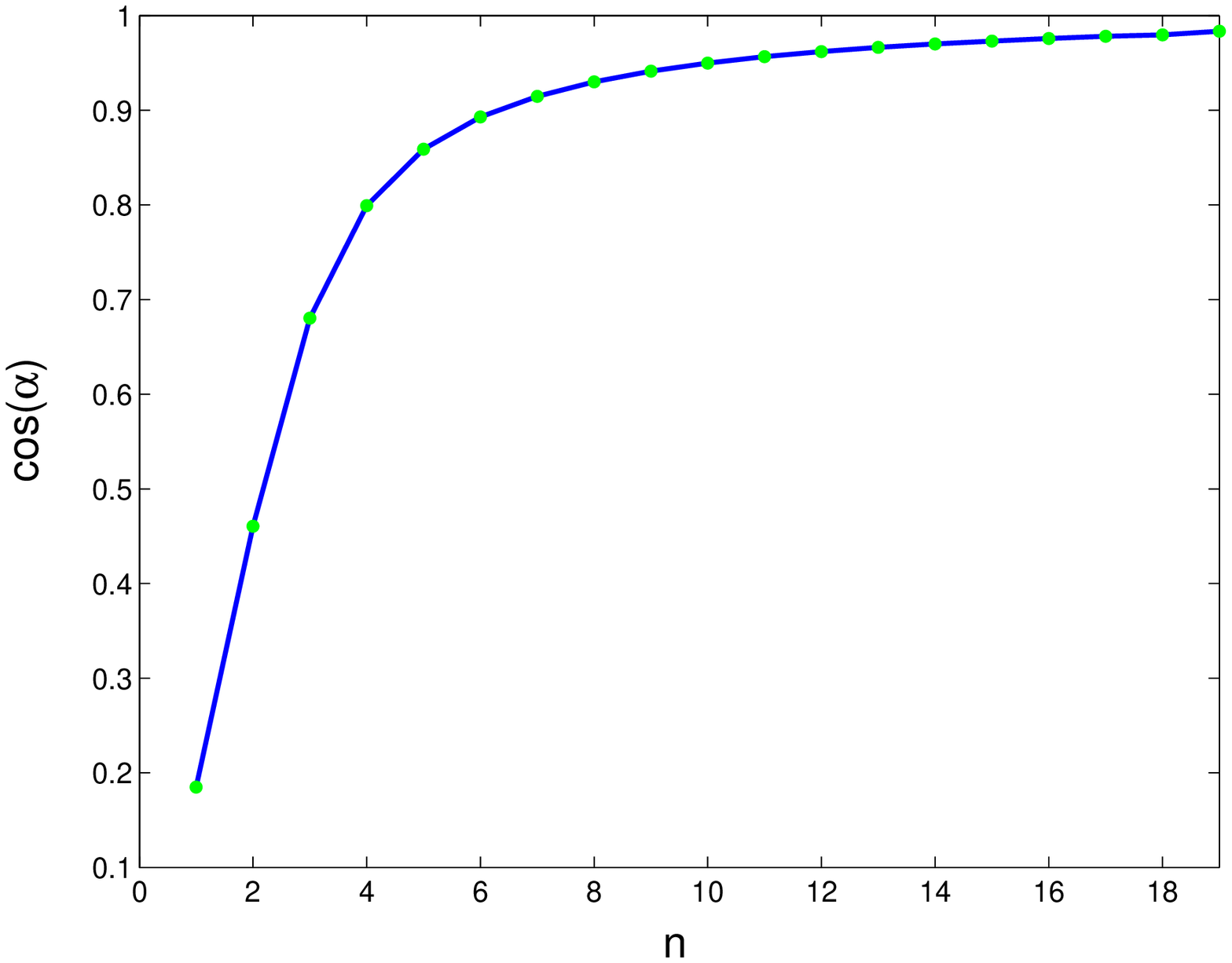}
\includegraphics[height= 6cm]{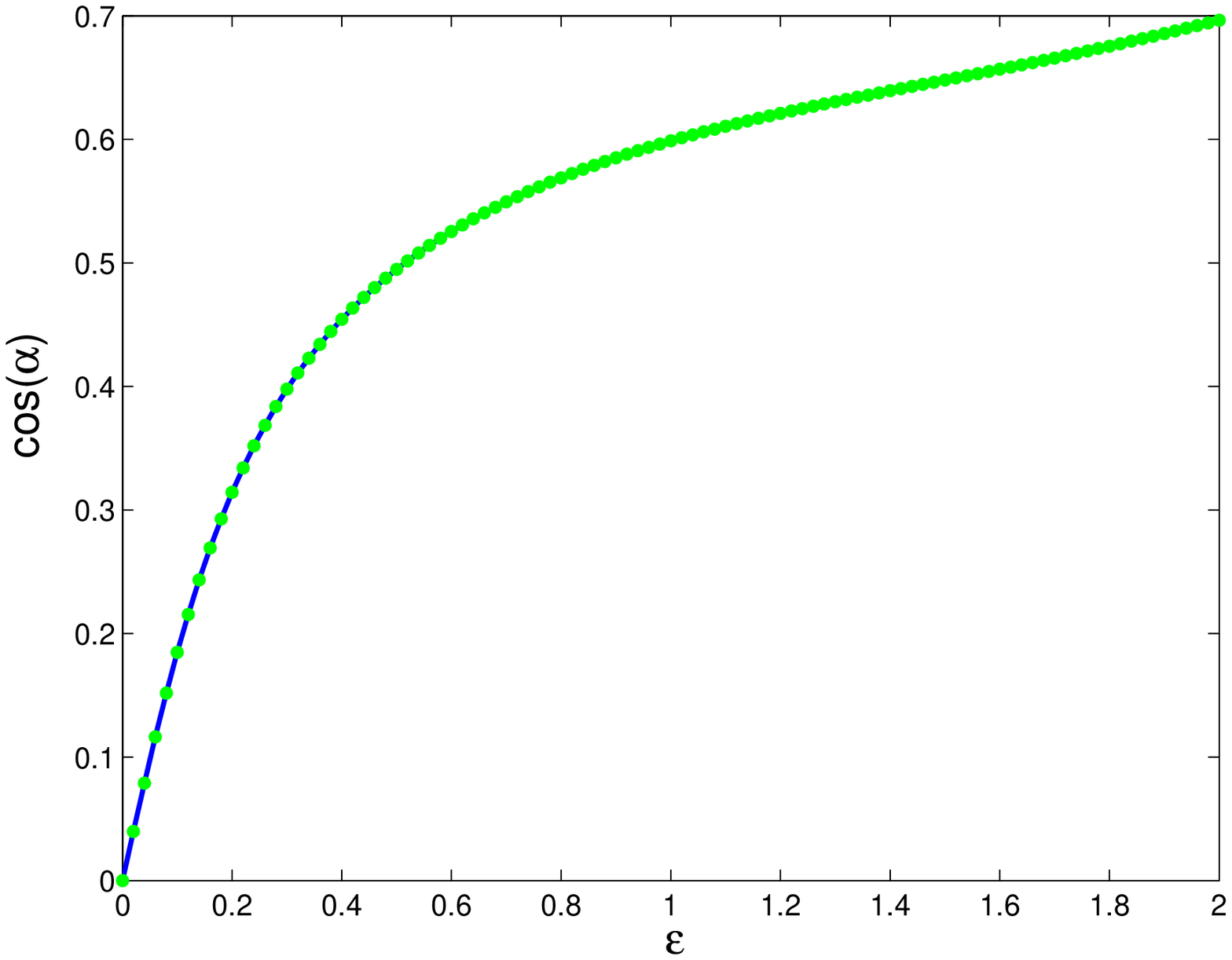}
\end{center}
\caption{Left: the values of $\cos(\widehat{f_n, f_{n+1}})$ for
the first 20 purely imaginary eigenvalues for $\epsilon = 0.1$.
Right: the values of $\cos(\widehat{f_1, f_2})$ versus
$\epsilon$.} \label{fig0}
\end{figure}

Figure \ref{fig0} (left) show the values of the cosine of the
angle (\ref{basis}) for the first $20$ purely imaginary
eigenvalues for $\epsilon = 0.1$. As we can see from the figure,
the angle between two eigenvectors tends to zero for larger
eigenvalues up to the numerical accuracy. Figure \ref{fig0}
(right) and Table 2 show that the angle drops to zero faster with
larger values of the parameter $\epsilon$.

\begin{center}
\begin{tabular}{|c|c|c|c|} \hline
eigenvectors  & $\epsilon = 0.1$ & $\epsilon = 0.3$ & $\epsilon = 0.5$ \\
  \hline
1-2 & 0.120166   &   0.325116  &   0.431987 \\
2-3 & 0.461330   &    0.716192 &    0.780641 \\
3-4 & 0.680709  &    0.838889  &   0.878055 \\
4-5 & 0.799235 &     0.890440  &   0.914622 \\
5-6& 0.858944  &    0.921498  &   0.940306 \\
6-7 & 0.892869  &    0.940395  &   0.955239 \\
7-8 & 0.914745  &    0.953124  &   0.965235 \\
8-9 & 0.930023 &     0.962120  &   0.972204 \\
9-10 & 0.941262 &     0.968732  &   0.977265 \\
10-11 & 0.949843  &    0.973741  &   0.981057 \\
11-12 & 0.956580  &    0.977629  &   0.983988 \\
12-13 & 0.961987   &   0.980702  &   0.986072 \\
13-14 & 0.966407   &   0.983297   &  0.989617 \\
14-15 & 0.970073  &    0.983459   &  0.990547 \\
15-16 & 0.973153  &    0.995335   &  0.999101 \\
16-17& 0.975764   &   0.998749   &  0.999601 \\
  \hline
\end{tabular}
\end{center}

{\bf Table 2:} Numerical values of $\cos(\widehat{f_n, f_{n+1}})$
for the first 16 purely imaginary eigenvalues for three values of
$\epsilon$.

The angle between two subsequent eigenvectors is closely related
to the condition number
\begin{equation}
\label{condition-number} \| P_n \| = \frac{\| f_n \| \|
f_n^*\|}{|(f_n, f_n^*)|} \equiv {\rm cond}(\lambda_n),
\end{equation}
which measures the norm of the spectral projections \cite{saad}.
By Lemma \ref{lemma-eigenvalues}(iii), the condition number is
infinite for multiple eigenvalues since $(f_n,f_n^*) = 0$. From
the point of numerical accuracy, the larger is the condition
number, the poorer is the structural stability of the numerically
obtained eigenvalues to the truncation and round-off errors.

\begin{figure}
\begin{center}
\includegraphics[height= 6cm]{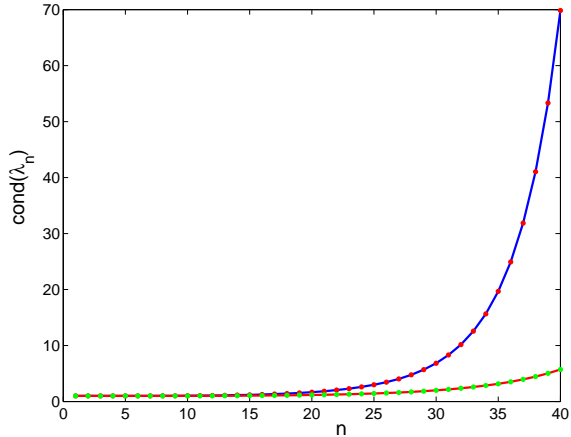}
\end{center}
\caption{The condition number for the first $40$ purely imaginary
eigenvalues for $\epsilon = 0.001$ (red) and $\epsilon = 0.002$
(blue).} \label{figcond}
\end{figure}

Figure \ref{figcond} shows the condition number
(\ref{condition-number}) computed for the first $40$ purely
imaginary eigenvalues for $\epsilon = 0.001$ and $\epsilon =
0.002$. We can see that the condition number grows for larger
eigenvalues which indicate their structural instability. Indeed,
starting with some number $n$, all eigenvalues are no longer
purely imaginary. The condition numbers become extremely large
with larger values of $\epsilon$.

We finally illustrate that all true eigenvalues of the spectral
problem (\ref{eigenvalue-problem}) are purely imaginary and
simple. To do so, we construct numerically the sign-definite
imaginary type function and obtain the interlacing property of
eigenvalues of the spectral problem (\ref{eigenvalue-problem}) for
two values $\epsilon = \epsilon_0$ and $\epsilon = \epsilon_1$,
where $|\epsilon_1 - \epsilon_0 |$ is small. We say that the
eigenvalues exhibit the interlacing property if there exists an
eigenvalue for $\epsilon = \epsilon_1$ between each pair of
eigenvalues for $\epsilon = \epsilon_0$ and vice verse.

A meromorphic function $G(\lambda)$ is called a sign-definite
imaginary type function if ${\rm Im} G(\lambda) \leq 0$ (${\rm Im}
G(\lambda) \geq 0$) on ${\rm Im}(\lambda) \leq 0$ (${\rm
Im}(\lambda) \geq 0$) \cite{atkinson}. We construct the
meromorphic function $G(\omega)$ in the form $G(\lambda) =
\frac{F_{\epsilon_0}(\lambda)}{F_{\epsilon_1}(\lambda)}$, where
$F_{\epsilon}(\lambda)$ is an analytical function of Corollary
\ref{corollary-accumulation}. The numerical approximation of the
meromorphic function $G(\lambda)$ is given by
$\widehat{G}(\lambda) =
\frac{\widehat{F}_{\epsilon_0}(\lambda)}{\widehat{F}_{\epsilon_1}(\lambda)}$.
According to Theorems II.2.1 - II.3.1 on p. 437-439 in
\cite{atkinson}, the function $\widehat{G}(\lambda)$ is a
meromorphic function of sign-definite imaginary type if and only
if it has the form $\widehat{G}(\lambda) =
\frac{P(\lambda)}{Q(\lambda)}$ where $P(\lambda)$ and $Q(\lambda)$
are polynomials with real coefficients, with real and simple
zeros, which are interlacing.

Table 3 shows this interlacing property of eigenvalues for
$\epsilon_0 = 0.48$ and $\epsilon_1 = 0.5$. The remainder term
$R_{\epsilon} = \frac{\| L f -\lambda f \|}{\| \lambda f \|}$
measures the numerical error of computations. We have also
computed numerically the values of $\widehat{G}(\lambda)$ on the
grid $ 0.1 < {\rm Im} \lambda < 100$ and $0.1 < {\rm Re} \lambda <
100$ with step size $0.1$ in both directions (not shown). Based on
the numerical data, we have confirmed that the function
$\widehat{G}(\lambda)$ does indeed belongs to the class of
sign-definite imaginary type functions while the eigenvalues $\{
\lambda_n \}_{n \in \mathbb{Z}}$ exhibit the interlacing property.
This computation gives a numerical verification that all
eigenvalues of the spectral problem (\ref{eigenvalue-problem}) are
simple and purely imaginary.

$$
\begin{array}{|c|c|c|c|}
\hline {\rm Im} \lambda_{\epsilon_0} & R_{\epsilon_0} &
{\rm Im} \lambda_{\epsilon_1} & R_{\epsilon_1}   \\
\hline
1.063112  &      2.32438e-10 &  1.068314  &    2.40727e-10    \\
2.970880  &     2.19667e-10  & 3.024428  &    2.25307e-10\\
5.414789  &       2.20243e-10&   5.542829  &    2.26833e-10\\
8.471510  &      2.0904e-10  &  8.693066  &    2.15721e-10\\
12.312548 &      2.00793e-10  & 12.665485 &    2.06007e-10\\
16.816692 &      1.97653e-10  & 17.327038 &    2.0288e-10\\
22.014084 &      1.96165e-10  & 22.711070 &    2.01973e-10\\
27.899896 &      1.95265e-10  & 28.812177 &    2.01571e-10\\
34.474785 &      1.95008e-10  & 35.631088 &    2.01904e-10\\
41.738699 &      1.95577e-10  & 43.167733 &    2.03125e-10\\
49.691673 &      1.96711e-10  & 51.422281 &    2.04763e-10\\
58.333258 &      1.97959e-10  & 60.391382 &    2.06229e-10\\
67.665387 &      1.99038e-10  & 70.140636 &    2.0725e-10\\
77.957871 &      1.99894e-10  & 79.828287 &    2.0782e-10\\
89.484519 &      2.65662e-10  & 91.544035 &    2.08206e-10\\
\hline
\end{array}%
$$
{\bf Table 3:} The interlacing property of the first $15$ purely
imaginary eigenvalues for $\epsilon = 0.48$ and $\epsilon = 0.5$.

\section{Conclusion}

We have proved that the operator $L$ associated with the heat
equation (\ref{heat}) admits a closure in $L^2([-\pi,\pi])$ with a
domain in $H^1_{\rm per}([-\pi,\pi])$ for $|\epsilon| < 2$. The
spectrum of $L$ consists of isolated eigenvalues of finite
multiplicities. Furthermore, we have proved with the assistance of
numerical computations that the set of eigenfunctions of the
spectral problem (\ref{eigenvalue-problem}) for isolated
eigenvalues is complete and minimal, but does not form a basis in
$H^1_{\rm per}([-\pi,\pi])$. By using the analytic function theory
and the truncated matrix eigenvalue problem, we have approximated
the eigenvalues numerically and showed that all eigenvalues of the
spectral problem (\ref{eigenvalue-problem}) are purely imaginary.

We conclude therefore that the spectral problem
(\ref{eigenvalue-problem}) is not useful for direct analysis of
well-posedness of the Cauchy problem for the heat equation
(\ref{heat}). This is very unusual compared to the case of a
standard heat equation
\begin{equation}
\label{standard-heat} \left\{ \begin{array}{ll} \dot{h} = -
h_{\theta} -
\epsilon h_{\theta \theta}, \quad t > 0,\\
h(0) = h_0, \end{array} \right.
\end{equation}
subject to the periodic boundary conditions on $\theta \in
[-\pi,\pi]$. By using the method of separation of variables and
the completeness of the Fourier series in $H^1_{\rm
per}([-\pi,\pi])$, one can easily prove that there exists a unique
solution of the heat equation (\ref{standard-heat}) for any $h_0
\in H^1_{\rm per}([-\pi,\pi])$ and any $\epsilon \in \mathbb{R}$
in the form
\begin{equation}
\label{decomposition-solution} h(t) = \sum_{n \in \mathbb{Z}} c_n
e^{\epsilon n^2 t} e^{i n (x-t)}, \quad t \geq 0,
\end{equation}
where the coefficients $\{ c_n \}_{n \in \mathbb{Z}}$ are found in
one and only one way from the initial data
\begin{equation}
\label{decomposition-initial} \forall h_0 \in H^1_{\rm
per}([-\pi,\pi]) : \quad h_0(\theta) = \sum_{n \in \mathbb{Z}} c_n
e^{i n x}.
\end{equation}

If $\epsilon > 0$, the Cauchy problem for the backward heat
equation (\ref{standard-heat}) is ill-posed. If $h_0$ is only in
$H^1_{\rm per}([-\pi,\pi])$, coefficients $\{ c_n \}_{n \in
\mathbb{Z}}$ decays to zero algebraically fast as $n \to \infty$
and the solution (\ref{decomposition-solution}) is singular for
any $t > 0$. Therefore, the solution $h(t)$ blows up in an
infinitesimally small time.

If $\epsilon < 0$, the Cauchy problem for the forward heat
equation (\ref{standard-heat}) is well-posed. In this case, the
solution $h(t)$ becomes analytic in $\theta \in [-\pi,\pi]$ for
any $t > 0$ even if $h_0$ is only in $H^1_{\rm per}([-\pi,\pi])$.
Therefore, there exists a constant $C > 0$ such that $\| h(t)
\|_{H^1_{\rm per}([-\pi,\pi])} \leq C \| h_0 \|_{H^1_{\rm
per}([-\pi,\pi])}$ for any $t \geq 0$.

Unlike this classical situation, the series of eigenfunctions of
the spectral problem (\ref{eigenvalue-problem}) are not applicable
to construct a solution of the periodic heat equation
(\ref{heat}), unless conditional convergence proposed in
\cite{Benilov2} can be adopted on a rigorous footing. Therefore,
the PDE analysis of the Cauchy problem for the periodic heat
equation (\ref{heat}) remains an open problem up to the date.

{\bf Note:} When the project was essentially complete, we became
aware of a preprint \cite{e-b-davies}, where similar results were
obtained. In particular, the author of \cite{e-b-davies} proves
that the spectral problem (\ref{eigenvalue-problem}) has no
essential spectrum and, assisted with the numerical computations,
illustrates that the eigenfunctions for all isolated eigenvalues
do not form a basis. The analysis of \cite{e-b-davies} is based on
the difference eigenvalue problem (\ref{difference-problem}),
which makes it different from our analysis.

{\bf Acknowledgement.} The authors thank E. Benilov for
formulation of the problem, V. Ivrii and V. Strauss for useful
discussions, and E.B. Davies for critical reading of our
manuscript. M.C. is supported by the NSERC Graduate Fellowship.
D.P. is supported by the Humboldt and EPSRC Research Fellowships.
The numerical work was made possible by the facilities of the
Shared Hierarchical Academic Research Computing Network
(SHARCNET).

\appendix

\section{Spectrum of the linear operator $L$ in weighted spaces}

The operator $L$ can be rewritten in the Sturm-Liouville symmetric
form
\begin{equation}
L = -\epsilon \left| \cot\left( \frac{\theta}{2}\right)
\right|^{1/\epsilon} L_0, \quad L_0 = \frac{d}{d \theta} \left(
\left| \tan\left(\frac{\theta}{2}\right) \right|^{1/\epsilon} \sin
\theta \frac{d}{d \theta} \right).
\end{equation}
Let $r(\theta) = |\tan(\theta/2)|^{1/\epsilon}$ be the weight of
the Sturm--Liouville spectral problem
\begin{equation}
-\epsilon L_0 f(\theta) = \lambda r(\theta) f(\theta),
\label{self-adjoint-L-0}
\end{equation}
acting on smooth functions $f(\theta)$ on $\theta \in [-\pi,\pi]$
in the weighted space $f \in L^2_{r}([-\pi, \pi])$.

\begin{proposition}
The operator $L_0$ admits a self-adjoint extension in
$L^2_r([-\pi,\pi])$ for $0 < \epsilon < 1$, such that the spectrum
of $L_0$ is purely discrete, consists of a set of simple real
eigenvalues $\{ \lambda_n \}_{n \in \mathbb{Z}}$ with $\lambda_0 =
0$, $\lambda_n = - \lambda_{-n}$, $\forall n \in \mathbb{N}$ and
$\lim\limits_{n \to \infty} \lambda_n = \infty$, and the
eigenfunction $f_n(\theta)$ for $\lambda_n > 0$ is identically
zero on $\theta \in [-\pi,0]$. \label{proposition-self-adjoint}
\end{proposition}

\begin{proof}
By Lemma \ref{lemma-analytic}, the eigenfunction $f(\theta)$ of
the spectral problem (\ref{self-adjoint-L-0}) for $0 < \epsilon <
1$ is in $L^2_r([-\pi,\pi])$ if and only if it is spanned by the
fundamental solutions $f_1(\theta)$, $f_2^+(\theta)$ and
$f_2^-(\theta)$, such that $f(\theta)$ is bounded at $\theta = 0$
and $\lim\limits_{\theta \to \pm \pi} f(\theta) = 0$. Let
$f(\theta)$ on $\pm \theta \in [0,\pi]$ be represented by
\begin{equation}
f(\theta) = \left[ \pm \cot \left( \frac{\theta}{2} \right)
\right]^{1/2\epsilon} g_{\pm}(x), \qquad \cos \theta = x,
\end{equation}
where $x \in [-1,1]$. Then, the spectral problem
(\ref{eigenvalue-problem}) is rewritten in the form
\begin{equation}
\label{second-order-g} -\epsilon \frac{d}{dx} \left[ (1-x^2)
\frac{dg_{\pm}}{dx} \right] + \frac{g_{\pm}(x)}{4 \epsilon
(1-x^2)} = \pm \frac{\lambda g_{\pm}(x)}{\sqrt{1-x^2}}.
\end{equation}
If $f(\theta)$ belongs to $L^2_r([-\pi,\pi])$ for $0 < \epsilon <
1$, then $g_{\pm}(x)$ belong to $L^2_{\rho}([-1,1])$ with the
weight function $\rho(x) = \frac{1}{\sqrt{1-x^2}}$. Since the
symmetric spectral problem (\ref{second-order-g}) is self-adjoint
in space $L_{\rho}^2([-1,1])$, its eigenvalues $\lambda$ are all
real-valued. By Lemma \ref{lemma-eigenvalues} and the
Sturm--Liouville theory \cite{CL}, the eigenvalues $\{ \lambda_n
\}_{n \in \mathbb{Z}}$ are all simple and symmetric with
$\lambda_0 = 0$ and $\lambda_n = - \lambda_{-n}$, $\forall n \in
\mathbb{N}$, while the sequence $\{ \lambda_n \}_{n \in
\mathbb{Z}}$ is unbounded with $\lim\limits_{n \to \infty}
\lambda_n = \infty$. By the standard Green's identity,
\begin{equation}
\label{identity-self-adjoint} \lambda \| g_+ \|^2_{L^2_{\rho}} =
\epsilon \int_{-1}^1 (1 - x^2) |g_+'(x)|^2 dx + \int_{-1}^1
\frac{|g_+(x)|^2 dx}{4 \epsilon (1-x^2)} > 0,
\end{equation}
the eigenvalues $\lambda_n$ of $g_+(x)$ are proved to be positive.
If $\lambda_n > 0$ is an eigenvalue for $g_+(x)$, then $-\lambda_n
< 0$ can not be an eigenvalue for $g_-(x)$, such that $g_-(x) = 0$
on $x \in [-1,1]$. Therefore, $f_n(\theta) = 0$ on $\theta \in
[-\pi,0]$ if $\lambda_n > 0$. By Lemma \ref{lemma-essential}, the
spectrum of $L$ is purely discrete for $|\epsilon | < 2$, such
that the spectrum of $L_0$ in $L^2_r([-\pi,\pi])$ is also purely
discrete.
\end{proof}

\begin{remark}
\label{remark-6} {\rm The operator $L_0$ admits the same
self-adjoint extension for any $\epsilon > 1$. However, this
extension is not unique. Indeed, if $\epsilon > 1$, the
eigenfunction $f(\theta)$ of the spectral problem
(\ref{self-adjoint-L-0}) may exist in $L^2_r([-\pi,\pi])$ even if
it is spanned by both fundamental solutions of Lemma
\ref{lemma-analytic} in the form
$$
f(\theta) = A f_1(\theta) + B f_2(\theta) = A^+ f_1^+(\theta) +
B^+ f_2^+(\theta) = A^- f_1^-(\theta) + B^- f_2^-(\theta).
$$
It follows from the Sturm--Liouville problem
(\ref{self-adjoint-L-0}) that
$$
\lambda \int_{-\pi}^{\pi} r(\theta) |f(\theta)|^2 d \theta = -
\epsilon \left| \tan \left( \frac{\theta}{2} \right)
\right|^{1/\epsilon} \sin(\theta) \bar{f}(\theta)
f^{\prime}(\theta) |_{\theta = -\pi}^{\theta = \pi} + \epsilon
\int_{-\pi}^{\pi} \left| \tan \left( \frac{\theta}{2} \right)
\right|^{1/\epsilon} \sin(\theta) |f^{\prime}(\theta)|^2 d\theta.
$$
The second integral is finite for $\epsilon > 1$ only if $B = 0$.
In this case, the first term is computed explicitly for $\epsilon
> 1$ as follows
\begin{equation}
2^{1/\epsilon} \left( \bar{A}^+ B^+ + \bar{A}^- B^- \right).
\label{term}
\end{equation}
This term is zero if $A^+ = A^- = 0$ for the self-adjoint
extension described in Proposition \ref{proposition-self-adjoint}.
However, this choice is not unique, e.g. the alternative pairing
$B^+ = B^- = 0$ can also be applied. }
\end{remark}

\section{Resonant poles of the Schr\"{o}dinger operators}

Eigenvalues of the operator $L$ coincide for $|\epsilon| >
\frac{1}{2}$ with resonant poles of the Schr\"{o}dinger operators.
To show this, we use the transformation
(\ref{Schrodinger-transformations}) on the intervals $\pm \theta
\in [0,\pi]$ and rewrite the spectral problem
(\ref{eigenvalue-problem}) as the uncoupled spectral problems
(\ref{Schrodinger-two-problems}) for $f(\theta) \equiv f_{\pm}(t)$
on $t \in \mathbb{R}$. Let us apply the normalization condition
$f(\pi) = f(-\pi) = 1$ for the eigenfunction $f(\theta)$ of the
spectral problem (\ref{eigenvalue-problem}) with $\lambda \notin
\mathbb{R}$. Indeed, if either $f(\pi) = 0$ or $f(-\pi) = 0$, then
the identity (\ref{identity-self-adjoint}) implies that $\lambda
\in \mathbb{R}$, but a real-valued $\lambda$ can not be an
eigenvalue of the spectral problem (\ref{eigenvalue-problem}) by
Lemma \ref{lemma-symmetry}. Therefore, eigenfunctions $f_{\pm}(t)$
of the uncoupled problems (\ref{Schrodinger-two-problems}) satisfy
the boundary conditions
\begin{equation}
\label{Schrodinger-boundary-conditions} \lim_{t \to -\infty}
f_{\pm}(t) = 1, \qquad \lim_{t \to \infty} f_{\pm}(t) = a^{\pm},
\end{equation}
where $a^{\pm}$ are uniquely defined. The function $f(\theta)$ on
$\theta \in [-\pi,\pi]$ constructed from $f_{\pm}(t)$ on $t \in
\mathbb{R}$ is continuous at $\theta = 0$ if $a^+ = a^-$.

Let $\epsilon > \frac{1}{2}$ and define $f_{\pm}(t) =
e^{\frac{t}{2 \epsilon}} g_{\pm}(t)$. The eigenfunctions
$g_{\pm}(t)$ satisfy the linear Schr\"{o}dinger equations
\begin{equation}
\label{Schrodinger-resonant-poles} \left( \frac{1}{4 \epsilon} -
\epsilon \partial_t^2 \right) g_{\pm} = \pm \lambda {\rm sech} t
\; g_{\pm},
\end{equation}
but the boundary conditions
(\ref{Schrodinger-boundary-conditions}) show that $g_{\pm} \notin
L^2(\mathbb{R})$ and $\lambda$ is not an eigenvalue. In fact, the
eigenfunctions $g_{\pm}(t)$ belong to the exponentially weighted
$L^2$-space, such that $\lambda$ is a resonance pole of the
Schr\"{o}dinger operators. Let us decompose the eigenfunctions by
$g_{\pm}(t) = e^{-\frac{t}{2 \epsilon}} + h_{\pm}(t)$. By the
theory of exponential asymptotics of solutions of the
Schr\"{o}dinger problems (\ref{Schrodinger-resonant-poles}), it
follows that $h_{\pm} \in L^2(\mathbb{R})$ if $\epsilon >
\frac{1}{2}$. Let $h_0(t) = e^{-\frac{t}{2 \epsilon}}\; {\rm sech}
t \in L^2(\mathbb{R})$ and define the linear inhomogeneous
problems for $h_{\pm}(t)$:
\begin{equation}
\label{Schrodinger-eigenvalues} S^{\pm}_{\epsilon}(\lambda)
h_{\pm} = \pm \lambda h_0(t), \qquad S^{\pm}_{\epsilon}(\lambda) =
\frac{1}{4 \epsilon} - \epsilon
\partial_t^2 \mp \lambda {\rm sech} t.
\end{equation}
The operator $S^{\pm}_{\epsilon}(\lambda)$ maps $H^2(\mathbb{R})$
to $L^2(\mathbb{R})$ and, if $\lambda \notin \mathbb{R}$, the
kernel of $S^{\pm}_{\epsilon}$ is empty. The boundary condition
$a^+ = a^-$ is then equivalent to the zeros of the function
\begin{equation}
H_{\epsilon}(\lambda) = \lambda \lim_{t \to \infty} e^{\frac{t}{2
\epsilon}} \left[ (S_{\epsilon}^+)^{-1}(\lambda) +
(S_{\epsilon}^-)^{-1}(-\lambda) \right] h_0(t).
\end{equation}
The function $H_{\epsilon}(\lambda)$ coincides (up to a
multiplicative constant) with the analytic function
$F_{\epsilon}(\lambda)$ introduced in Corollary
\ref{corollary-accumulation} for $|\epsilon| < 2$. Therefore,
$H_{\epsilon}(\lambda)$ represents a continuation of
$F_{\epsilon}(\lambda)$ from the domain $|\epsilon| < 2$ to the
domain $|\epsilon | > \frac{1}{2}$. The function
$H_{\epsilon}(\lambda)$ is analytic in $\lambda \in \mathbb{C}$
and its roots give isolated eigenvalues of the spectral problem
(\ref{eigenvalue-problem}) with the account of their multiplicity.

The function $H_{\epsilon}(\lambda)$ can be simplified for
$\lambda \in i \mathbb{R}$. Let $\lambda = i \omega \in i
\mathbb{R}$ and represent $h_{\pm} = F(t) \pm i G(t)$, where
\begin{equation}
L_{\epsilon} F = - \omega \; {\rm sech} t \; G, \qquad
L_{\epsilon} G = \omega \; {\rm sech} t \; F + \omega h_0(t),
\qquad L_{\epsilon} = \frac{1}{4 \epsilon} - \epsilon
\partial_t^2.
\end{equation}
Therefore, we can define a real-valued function
$\tilde{H}_{\epsilon}(\omega) = H_{\epsilon}(i \omega)$ on $\omega
\in \mathbb{R}$ given by
\begin{equation}
\tilde{H}_{\epsilon}(\omega) = \omega \lim_{t \to \infty}
e^{\frac{t}{2 \epsilon}} \left( L_{\epsilon} + \omega^2 \; {\rm
sech} t \; L_{\epsilon}^{-1} \; {\rm sech} t \right)^{-1} h_0(t).
\end{equation}
By multiplying the linear inhomogeneous equation
$$
\left( L_{\epsilon} + \omega^2 \; {\rm sech} t \;
L_{\epsilon}^{-1} \; {\rm sech} t \right) G = \omega h_0(t),
$$
by $e^{\frac{t}{2 \epsilon}}$ and integrating on $t \in
\mathbb{R}$ by parts, the function $\tilde{H}_{\epsilon}(\omega)$
can be represented in the integral form
$$
\tilde{H}_{\epsilon}(\omega) = \omega \int_{-\infty}^{\infty} {\rm
sech} t dt - \omega^2 \int_{-\infty}^{\infty} e^{\frac{t}{2
\epsilon}} {\rm sech} t \; L_{\epsilon}^{-1} \; {\rm sech} t G(t)
dt.
$$
Let $\tilde{h}_0(t) = {\rm sech} t \; L_{\epsilon}^{-1} \; {\rm
sech} t e^{\frac{t}{2\epsilon}}$. Then,
$$
\tilde{H}_{\epsilon}(\omega) = \pi \omega - \omega^2
\int_{-\infty}^{\infty} \tilde{h}_0(t) G(t) dt.
$$
This is yet another form of the real analytic function, zeros of
which are equivalent to purely imaginary eigenvalues of the
spectral problem (\ref{eigenvalue-problem}) for $|\epsilon|
> \frac{1}{2}$.

\end{document}